\documentclass[preprint, 12pt, amsmath,amssymb, aps, prd]{revtex4-1}
\usepackage[utf8]{inputenc}
\usepackage{hyperref}
\usepackage{slashed}
\usepackage{color}
\usepackage{graphicx}
\graphicspath{{pics/}}
\allowdisplaybreaks[1]

\begin{document}
\title{Photons production in heavy-ion collisions as a signal of deconfinement phase}

\author{Sergei Nedelko}
\email{nedelko@theor.jinr.ru}
\author{Aleksei Nikolskii}
\email{alexn@theor.jinr.ru}
\affiliation{Bogoliubov Laboratory of Theoretical Physics, JINR, 141980 Dubna, Russia}

\begin{abstract}

The photon production  due to  conversion  of two gluons into a photon,  $gg\rightarrow\gamma$,  in the presence of the background gauge fields is studied  within the specific mean-field approach to QCD vacuum. In this approach,  mean field in the confinement phase is represented by the statistical ensemble of almost everywhere homogeneous abelian (anti-)self-dual gluon configurations,  while the deconfined phase can be characterized by the purely chromomagnetic fields. The probability of gluon conversion of two gluons into a photon vanishes in the confinement phase due to the randomness of the background field configurations. The anisotropic strong electromagnetic field, generated in the collision  of relativistic heavy ions, serves as a catalyst for deconfinement  with the appearance of an anisotropic purely chromomagnetic mean field. Respectively, deconfined phase is characterized by nonzero  probability of the conversion of two gluons into a photon  with strongly  anisotropic  angular distribution. 

\end{abstract}

\maketitle

\section{Introduction}

Extremely strong electromagnetic fields  generated during the  collisions of relativistic heavy ions (Refs.~\cite{Skokov:2009qp,Voronyuk:2011jd}) may lead to numerous  physical effects, proposed in the literature~\cite{Bzdak:2012fr,Shovkovy:2012zn,Tuchin:2013ie,Ayala:2017vex,d2021phase}. The generated magnetic field has a preferred spatial orientation which  is manifested through the angular anisotropies in various observables.  Besides the direct effects of quark field interaction with the electromagnetic field there can be even more  vigorous effects related to a kind of polarization of the QCD vacuum caused by the effective interaction of gluon and  electromagnetic fields via their coupling to the quark fields, which  are known to play the catalyzing role for deconfinement (Refs.~\cite{Galilo:2011nh, Nedelko:2014sla, DElia:2012ifm, Bali:2013esa, DElia:2021tfb, Bonati:2016kxj, Fukushima:2012xw,Fukushima:2012xw}). A plausible interpretation of the mechanism of the catalysis was offered in the literature~\cite{Galilo:2011nh, Nedelko:2014sla,Ozaki:2013sfa} within the mean-field approach to nonperturbative QCD vacuum. The  main observation of these papers is that in the presence of an external magnetic field  there exists a global minimum of the one-loop quark contribution to the effective potential of QCD corresponding to a purely chromomagnetic ($F^a_{\mu\nu}\tilde F^a_{\mu\nu}=0$) gluon field. In contrast to the abelian (anti-)self-dual homogeneous gluon field, which is a  plausible candidate for a global minimum of the effective QCD potential in the absence of external electromagnetic fields (e.g., see Refs.~\cite{Leutwyler:1980ma,Nedelko:2020bba}), the chromomagnetic gluon field does not support confinement, since the color charged quasiparticles do exist and can move along the direction of the chromomagnetic field, which in turn coincides with the direction of magnetic field (Ref.~\cite{Nedelko:2014sla}). Such an interpretation appears to be consistent with the lattice QCD studies (Refs.~\cite{Bali:2013esa,DElia:2012ifm,Bonati:2018uwh}). 
 
It should be noted that according to the mean-field interpretation an anisotropic background chromomagnetic field can exist as long as the deconfinement phase occurs, unlike the short-lived extreme magnetic field which simply triggers the anisotropy (for details see Ref.~\cite{Nedelko:2014sla}). The presence of the  chromomagnetic background gauge field violates the conditions of the Furry theorem, and thus  leads to the possibility of conversion of a pair of gluons into a photon through a quark loop, which can be seen  as an important mechanism for the generation of the direct photons in the deconfinement phase, similarly to the case of magnetic background discussed in Refs.~\cite{Ayala:2017vex,Ayala:2019jey,Adler:1970gg,Papanyan:1971cv,Baier:1974hn}. 
 
An abnormally high  photon yields and degree of azimuthal anisotropy were observed in experiments by ALICE and PHENIX collaborations (Refs.~\cite{PHENIX:2011oxq,PHENIX:2014nkk,ALICE:2015xmh}). This effect is known as direct photon flow puzzle. Various phenomenological explanations of this phenomenon were proposed over the years in the literature~\cite{Shen:2016egw,Paquet:2015lta,Chatterjee:2007xk,vanHees:2011vb,Bzdak:2012fr,Liu:2012ax,Linnyk:2013hta,Zakharov:2016mmc,Zakharov:2017cul,Vujanovic:2017wtw,Goloviznin:2012dy,Goloviznin:2018mwb}. The low momentum $q_{\perp}$ part  of the photon spectra is well characterized by their inverse logarithmic temperature slope $T_{\rm eff}$. The PHENIX collaboration found $T_{\rm eff}=239$ MeV in 0-20 \% Au+Au collisions at $\sqrt{s_{NN}}=200$ GeV (Ref.~\cite{PHENIX:2014nkk}) and the ALICE collaboration found $T_{\rm eff}=297$ MeV in 0-20 \% Pb+Pb collisions at $\sqrt{s_{NN}}=2.76$ TeV (Ref.~\cite{ALICE:2015xmh}). Calculations based on the hydrodynamic models (Refs.~\cite{Paquet:2015lta,Chatterjee:2007xk,vanHees:2011vb,Liu:2012ax}) lead to agreement with the experimental photon spectrum for the range $q_{\perp} > 1$ GeV. A promising approach  to explain the direct photon spectrum are the studies  within  parton-hadron-string dynamics (PHSD) based on the microscopic transport calculations (Ref.~\cite{Linnyk:2013hta}). Also, the photon production in quark-gluon plasma (QGP) due to the bremsstrahlung was estimated in Refs.~\cite{Zakharov:2016mmc,Zakharov:2017cul,Goloviznin:2012dy,Goloviznin:2018mwb}. A mechanism based on Furry theorem violation by the electromagnetic fields generated in  heavy-ion collisions has been proposed relatively recently in Refs.~\cite{Ayala:2017vex,Ayala:2019jey}. In general, the situation with the description of the direct photon spectrum is improving, but some tension remains between theory and experiment (Ref.~\cite{Shen:2016odt}). 

\begin{figure}[!h]
\center{\includegraphics[scale=0.7]{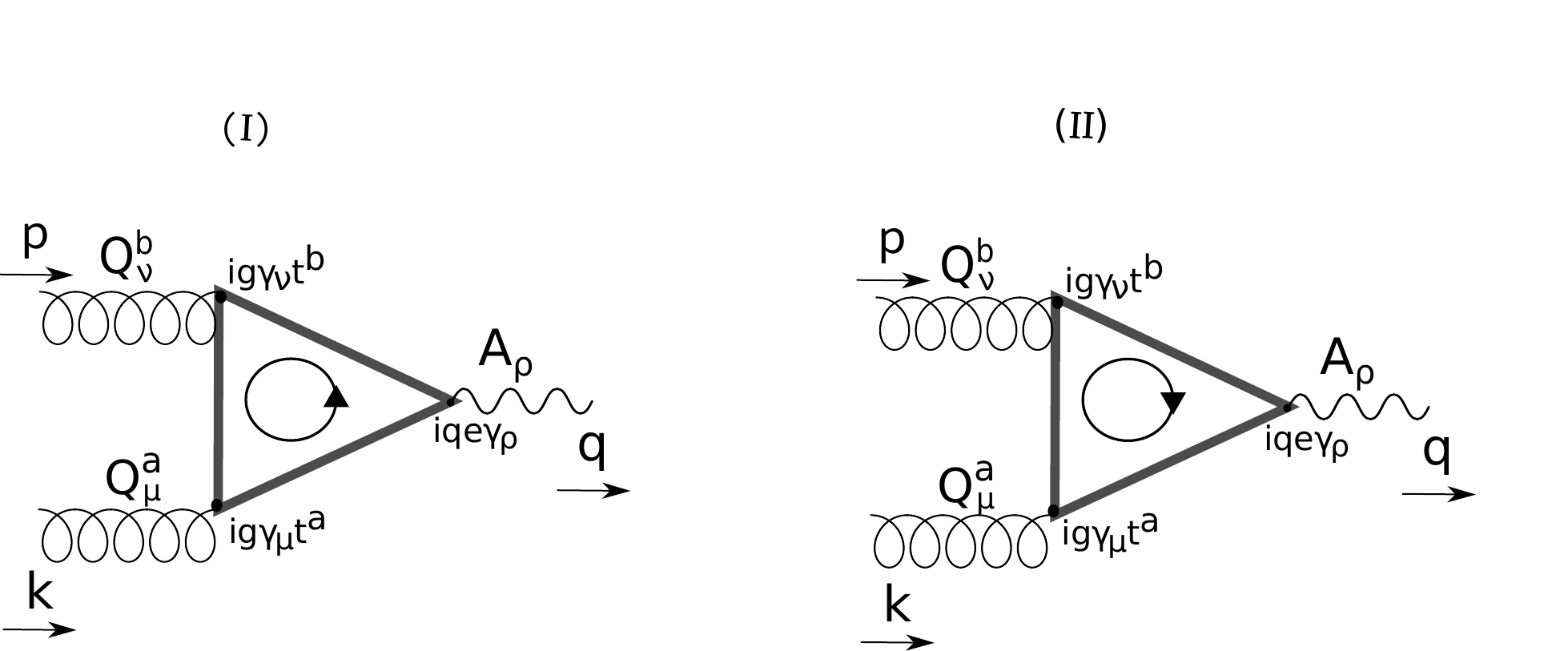}}
\caption{The diagrams for the process $gg \rightarrow \gamma$. Here $p$, $k$ - are  momenta of the gluons, $q$ is the photon momentum. The arrows inside loop indicate the direction of loop momentum.}
\label{th_loop}
\end{figure}

In this paper we calculate the contribution of the quark loop shown in Fig.~\ref{th_loop} to the process $ gg \rightarrow \gamma $ for  two regimes: in the presence of confining vacuum mean-field, represented by the statistical  ensemble of the almost everywhere homogeneous (anti-)self-dual abelian gluon  fields, and for the case of anisotropic chromomagnetic field characteristic of the deconfinement regime. In the confinement phase, due to the random nature of the mean-field,  this contribution vanishes on average. In deconfinement regime the direction of the chromomagnetic field is correlated with the direction of the short-lived generated magnetic field, the conditions of the Furry theorem are thus  violated, and the diagram in Fig.~\ref{th_loop} gives a nonzero contribution. Due to its long-lived nature and high field strength, defined as a matter of fact by the  value of the scalar gluon condensate $\langle g^2F^2\rangle$, the chromomagnetic background has clearly promising potential for explaining both  puzzling features of direct photon measurements, spectra and anisotropy, simultaneously. The  estimate elaborated in line with Refs.~\cite{Ayala:2017vex,Ayala:2019jey} indicates rather strong effect of  conversion of the gluons to a photon in the deconfinement phase. The appearance of an additional photon source due to the process $ gg \rightarrow \gamma $ may be seen as a signal for deconfinement. 

Studies of the present paper have to be considered in the context of the domain model of QCD vacuum and hadronization (for details see Refs.~\cite{Leutwyler:1980ma,Galilo:2011nh, Nedelko:2014sla,Nedelko:2016gdk,Nedelko:2016vpj,Nedelko:2020bba}). The model is based on the vacuum mean-field represented by the ensemble of domain-structured configurations of almost everywhere homogeneous abelian (anti-)self-dual gluon field which is treated nonperturbatively. This mean field provides simultaneously  confinement of static and dynamic quarks - the area law for the Wilson loop and the absence of poles in the quark propagator in the complex momentum plane respectively, as well as flavour chiral symmetry breaking and the resolution of the $ U_{\rm A}(1)$ problem. Upon bosonization, this mean-field approach successfully describes   the masses of light, heavy-light mesons and heavy quarkonia, including their excited states, as well as the decay constants and form factors (Refs.~\cite{Nedelko:2016gdk,Nedelko:2016vpj}). The mean field does not affect the UV-behaviour of quark, gluon and ghost propagators, but otherwise strongly modifies the propagators, whose form is overall consistent with the results of the functional renormalization group and lattice QCD (for complete analysis see Ref.~\cite{Nedelko:2016gdk}). 

\section{The confinement phase}
In this section, we consider the amplitude for the process $gg\rightarrow\gamma$ \textit{via} a quark loop (Fig.~\ref{th_loop})  in the presence of the  homogeneous abelian (anti-)self-dual gluon field: 
\begin{gather}
\hat{B}_\mu=\frac{1}{2}\hat{B}_{\mu\nu}x_\nu,~\hat{B}_{\mu\nu}=\hat{n}B_{\mu\nu},~\hat{n}=t^8=\frac{\lambda^8}{2}, \nonumber \\
\widetilde{B}_{\mu\nu}=\frac{1}{2}\epsilon_{\mu\nu\alpha\beta}B_{\alpha\beta}=\pm B_{\mu\nu},~ \hat{B}_{\rho\mu}\hat{B}_{\rho\nu}=4v^2B^2\delta_{\mu\nu}, \nonumber \\
\label{self-field}
\hat{f}_{\alpha\beta}=\frac{\hat{n}}{2vB}B_{\alpha\beta},~v=\text{diag}\left(\frac{1}{6},\frac{1}{6},\frac{1}{3}\right),~\hat{f}^{ik}_{\mu\alpha}\hat{f}^{kj}_{\nu\alpha}=\delta^{ij}\delta_{\mu\nu}, 
\end{gather}
where  $\lambda^8$ is the  Gell-Mann matrix.  Field strength $B$ sets the scale related to the value of the scalar gluon condensate. 

The propagator of the quark field with mass $m_f$ in the presence of the (anti-)self-dual field determined in Eq.~\eqref{self-field} has the form (Ref.~\cite{{Nedelko:2016gdk}})
\begin{eqnarray}
\label{self-propagator}
S_f(x,y)&=&\exp\left(\frac{i}{2}x_\mu \hat{B}_{\mu\nu} y_\nu\right)H_f(x-y), \\
H_f(z)&=&\frac{vB}{8\pi^2}\int_0^1 \frac{ds}{s^2} \exp\left(-\frac{vB}{2s}z^2\right)\left(\frac{1-s}{1+s}\right)^{\frac{m_f^2}{4vB}} \nonumber \\
&\times&\left[-i\frac{vB}{s}z_\mu \left(\gamma_\mu \pm is\hat{f}_{\mu\nu}\gamma_\nu\gamma_5 \right) +m_f\left(P_\pm+\frac{1+s^2}{1-s^2}P_{\mp}+\frac{i}{2}\gamma_\mu \hat{f}_{\mu\nu} \gamma_\nu \frac{s}{1-s^2}\right)\right], \nonumber \\
\nonumber 
\end{eqnarray} 
where $z=x-y$, $P_\pm=(1\pm \gamma_5)\big/2$ and the anti-Hermitean representation of the Dirac matrices in Euclidean space-time is used. Sign ``$\pm$"  corresponds to (anti-)self-duality of the background field in Eq.~\eqref{self-field}. The Fourier transform of the translation invariant part $H_f$ of the propagator is an entire analytical function in the complex momentum plane. It approaches the limit of the standard free Dirac propagator at large Euclidean momentum $p^2\gg B$. 

The amplitudes for diagrams $(I)$ and $(II)$ in Fig.~\ref{th_loop} take the form
\begin{gather}
\label{Ma1qcd}
M^{(I)}=ieg^2\int d^{4}xd^{4}yd^{4}ze^{-i(px+ky-qz)}\left\langle\mathrm{Tr}\left[\gamma_{\nu}t^{b}S(x,z)Q\gamma_{\rho}S(z,y)\gamma_{\mu}t^{a}S(y,x)\right]\right\rangle\epsilon^{a}_{\mu}(k)\epsilon^{b}_{\nu}(p)\epsilon_{\rho}(q),
 \nonumber\\
\label{Mb1qcd}
M^{(II)}=ieg^2\int d^{4}xd^{4}yd^{4}ze^{-i(px+ky-qz)}\left\langle\mathrm{Tr}\left[S(x,y)\gamma_{\mu}t^{a}S(y,z)Q\gamma_{\rho}S(z,x)\gamma_{\nu}t^{b}\right]\right\rangle\epsilon^{a}_{\mu}(k)\epsilon^{b}_{\nu}(p)\epsilon_{\rho}(q), 
\nonumber
\end{gather}
here $g$ is the strong coupling constant, $e$ is the electron charge, $Q$ is a diagonal matrix of the fractions of electric charges of  quarks with flavor $f$, vectors $\epsilon$ define the polarization of the gluons and photon. Tr denotes trace of the color, Dirac and flavor matrices, and $\langle\dots\rangle$ denotes averaging of the amplitude over different random configurations  of the background  field: (anti-)self-duality and spatial orientation. In particular, integration over spatial orientations of the background field is given by the master formula (Ref.~\cite{Nedelko:2016vpj})
\begin{gather}
\label{av-field}
\langle\exp(if_{\mu\nu}J_{\mu\nu})\rangle=\frac{\sin\sqrt{2\left(J_{\mu\nu}J_{\mu\nu}\pm J_{\mu\nu}\widetilde{J}_{\mu\nu}\right)}}{\sqrt{2\left(J_{\mu\nu}J_{\mu\nu}\pm J_{\mu\nu}\widetilde{J}_{\mu\nu}\right)}},
\end{gather}
where $J_{\mu\nu}$ is an arbitrary antisymmetric tensor, $\widetilde{J}_{\mu\nu}=\frac{1}{2}\epsilon_{\mu\nu\alpha\beta}J_{\alpha\beta}$. 
 
Taking into account Eq.~\eqref{self-propagator} one may integrate over one of the spatial coordinates and arrive at the representation  
\begin{eqnarray}
\label{Ma2qcd}
M^{(I)}&=&ieg^2(2\pi)^{4}\delta^{(4)}(p+k-q)\int d^{4}xd^{4}y \ e^{-i(px+ky)} \nonumber \\
&\times & \left\langle\text{Tr}\left[e^{-ivBy^{\mu}\hat{f}_{\mu\nu}x^{\nu}}\gamma_{\nu}t^{b}H(x) Q\gamma_{\rho}H(-y)~\gamma_{\mu}t^{a}H(y-x)\right]\right\rangle\epsilon^{a}_{\mu}(k)\epsilon^{b}_{\nu}(p)\epsilon_{\rho}(q), \\
\label{Mb2qcd}
M^{(II)}&=&ieg^2(2\pi)^{4}\delta^{(4)}(p+k-q)\int d^{4}xd^{4}y \ e^{-i(px+ky)} \nonumber \\ 
&\times &\left \langle\mathrm{Tr}\left[e^{-ivBx^{\mu}\hat{f}_{\mu\nu}y^{\nu}}H(x-y)\gamma_{\mu}t^{a}H(y)Q\gamma_{\rho}H(-x)\gamma_{\nu}t^{b}\right]\right\rangle \epsilon^{a}_{\mu}(k)\epsilon^{b}_{\nu}(p)\epsilon_{\rho}(q),
\end{eqnarray} 
The terms in the amplitudes $M^{(I)}$ and $M^{(II)}$ with odd  powers of field strength tensor $\hat{f}_{\mu\nu}$ violate the conditions of the Furry theorem. It becomes explicit by substitution of quark propagator $H_f(z)$ in Eq.~\eqref{Ma2qcd} and Eq.~\eqref{Mb2qcd} resulting in the following representation
\begin{gather}
\label{Ma2qcdexpand1}
M^{(I)}=ieg^2q_f(2\pi)^{4}\delta^{(4)}(p+k-q) \left(\frac{vB}{8\pi^2}\right)^3 \int_0^1 \int_0^1 \int_0^1 \frac{ds_1}{s_1^2} \frac{ds_2}{s_2^2} \frac{ds_3}{s_3^2} \frac{(-ivB)^3}{s_1s_2s_3} \nonumber \\\left(\frac{1-s_1}{1+s_1}\right)^{\frac{m^2_f}{4vB}} \left(\frac{1-s_2}{1+s_2}\right)^{\frac{m^2_f}{4vB}} \left(\frac{1-s_3}{1+s_3}\right)^{\frac{m^2_f}{4vB}} \int d^{4}xd^{4}y \ e^{-i(px+ky)} \nonumber \\
\times \Big\langle\text{Tr}\Big[e^{-ivBy^{\mu}\hat{f}_{\mu\nu}x^{\nu} -\frac{v}{2s_1}x^2 -\frac{v}{2s_2}y^2 - \frac{v}{2s_3}(y-x)^2} \nonumber \\ 
\hat{f}_{\alpha \omega}  \hat{f}_{\beta \chi}  \hat{f}_{\lambda \eta} \left( \pm s_1s_2s_3 ~ x_\alpha x_\beta y_\lambda  \gamma_5\gamma_\nu\gamma_{\omega}\gamma_\rho\gamma_{\eta}\gamma_\mu\gamma_{\chi}   
\mp s_1s_2s_3 ~ x_\alpha y_\beta y_\lambda \gamma_5\gamma_\nu\gamma_{\omega}\gamma_\rho\gamma_{\eta}\gamma_\mu\gamma_{\chi} \right) \nonumber \\
\hat{f}_{\alpha \eta} \hat{f}_{\beta \omega} \left(-i s_2s_3 x_\alpha y_\beta x_\lambda \gamma_\nu\gamma_\lambda\gamma_\rho\gamma_{\omega}\gamma_\mu\gamma_{\eta} 
+i s_2s_3 y_\alpha y_\beta x_\lambda \gamma_\nu\gamma_\lambda\gamma_\rho\gamma_{\omega}\gamma_\mu\gamma_{\eta} \right. 
\nonumber \\
\left. -i s_1s_3 x_\alpha y_\beta x_\lambda \gamma_\nu\gamma_{\omega}\gamma_\rho\gamma_{\lambda}\gamma_\mu\gamma_{\eta} 
+i s_1s_3 x_\alpha y_\beta x_\lambda \gamma_\nu\gamma_{\omega}\gamma_\rho\gamma_{\beta}\gamma_\mu\gamma_{\eta} \right. 
\nonumber \\
\left. -i s_1s_2 x_\alpha y_\beta x_\lambda \gamma_\nu\gamma_{\omega}\gamma_\rho\gamma_{\eta}\gamma_\mu\gamma_{\alpha} 
+i s_1s_2 y_\alpha y_\beta x_\lambda \gamma_\nu\gamma_{\omega}\gamma_\rho\gamma_{\eta}\gamma_\mu\gamma_{\alpha} \right) 
\nonumber \\
\hat{f}_{\alpha \omega} \left( \mp s_3 x_\alpha x_\beta y_\lambda  \gamma_5\gamma_\nu\gamma_\beta\gamma_\rho\gamma_\lambda\gamma_\mu
\gamma_{\omega} 
\pm s_3 y_\alpha x_\beta y_\lambda  \gamma_5\gamma_\nu\gamma_\beta\gamma_\rho\gamma_\lambda\gamma_\mu
\gamma_{\omega} 
\mp s_2 y_\alpha x_\beta x_\lambda  \gamma_5\gamma_\nu\gamma_\beta\gamma_\rho\gamma_{\omega}\gamma_\mu
\gamma_{\lambda} \right. 
\nonumber \\ 
\left. \pm s_2 y_\alpha x_\beta y_\lambda  \gamma_5\gamma_\nu\gamma_\beta\gamma_\rho\gamma_{\omega}\gamma_\mu
\gamma_{\lambda} 
\mp s_1 x_\alpha y_\beta x_\lambda  \gamma_5\gamma_\nu\gamma_{\omega}\gamma_\rho\gamma_{\beta}\gamma_\mu
\gamma_{\lambda} 
\pm s_1 x_\alpha y_\beta y_\lambda  \gamma_5\gamma_\nu\gamma_{\omega}\gamma_\rho\gamma_{\beta}\gamma_\mu
\gamma_{\lambda} \right) \nonumber \\ 
+ i x_\alpha y_\beta x_\lambda \gamma_\nu\gamma_{\alpha}\gamma_\rho\gamma_{\beta}\gamma_\lambda 
- i x_\alpha y_\beta y_\lambda \gamma_\nu\gamma_{\alpha}\gamma_\rho\gamma_{\beta}\gamma_\lambda \Big] \Big\rangle ~\epsilon^{a}_{\mu}(k)\epsilon^{b}_{\nu}(p)\epsilon_{\rho}(q), \nonumber 
\end{gather}

\begin{gather}
\label{Ma2qcdexpand2}
M^{(II)}=ieg^2q_f(2\pi)^{4}\delta^{(4)}(p+k-q) \left(\frac{vB}{8\pi^2}\right)^3 \int_0^1 \int_0^1 \int_0^1 \frac{ds_1}{s_1^2} \frac{ds_2}{s_2^2} \frac{ds_3}{s_3^2} \frac{(-ivB)^3}{s_1s_2s_3} \nonumber \\\left(\frac{1-s_1}{1+s_1}\right)^{\frac{m^2_f}{4vB}} \left(\frac{1-s_2}{1+s_2}\right)^{\frac{m^2_f}{4vB}} \left(\frac{1-s_3}{1+s_3}\right)^{\frac{m^2_f}{4vB}} \int d^{4}xd^{4}y \ e^{-i(px+ky)} \nonumber \\
\times \Big\langle\text{Tr}\Big[e^{-ivBx^{\mu}\hat{f}_{\mu\nu}y^{\nu} -\frac{v}{2s_1}(x-y)^2 -\frac{v}{2s_2}y^2 - \frac{v}{2s_3}x^2} \nonumber \\ 
\hat{f}_{\alpha \omega}  \hat{f}_{\beta \chi}  \hat{f}_{\lambda \eta} \left( \pm s_1s_2s_3 ~ x_\alpha x_\beta y_\lambda  \gamma_5\gamma_\nu\gamma_{\omega}\gamma_\rho\gamma_{\eta}\gamma_\mu\gamma_{\chi}   
\mp s_1s_2s_3 ~ x_\alpha y_\beta y_\lambda \gamma_5\gamma_\nu\gamma_{\omega}\gamma_\rho\gamma_{\eta}\gamma_\mu\gamma_{\chi} \right) \nonumber \\
-\hat{f}_{\alpha \eta} \hat{f}_{\beta \omega} \left(-i s_2s_3 x_\alpha y_\beta x_\lambda \gamma_\nu\gamma_\lambda\gamma_\rho\gamma_{\omega}\gamma_\mu\gamma_{\eta} 
+i s_2s_3 y_\alpha y_\beta x_\lambda \gamma_\nu\gamma_\lambda\gamma_\rho\gamma_{\omega}\gamma_\mu\gamma_{\eta} \right. 
\nonumber \\
\left. -i s_1s_3 x_\alpha y_\beta x_\lambda \gamma_\nu\gamma_{\omega}\gamma_\rho\gamma_{\lambda}\gamma_\mu\gamma_{\eta} 
+i s_1s_3 x_\alpha y_\beta x_\lambda \gamma_\nu\gamma_{\omega}\gamma_\rho\gamma_{\beta}\gamma_\mu\gamma_{\eta} \right. 
\nonumber \\
\left. -i s_1s_2 x_\alpha y_\beta x_\lambda \gamma_\nu\gamma_{\omega}\gamma_\rho\gamma_{\eta}\gamma_\mu\gamma_{\alpha} 
+i s_1s_2 y_\alpha y_\beta x_\lambda \gamma_\nu\gamma_{\omega}\gamma_\rho\gamma_{\eta}\gamma_\mu\gamma_{\alpha} \right) 
\nonumber \\
\hat{f}_{\alpha \omega} \left( \mp s_3 x_\alpha x_\beta y_\lambda  \gamma_5\gamma_\nu\gamma_\beta\gamma_\rho\gamma_\lambda\gamma_\mu
\gamma_{\omega} 
\pm s_3 y_\alpha x_\beta y_\lambda  \gamma_5\gamma_\nu\gamma_\beta\gamma_\rho\gamma_\lambda\gamma_\mu
\gamma_{\omega} 
\mp s_2 y_\alpha x_\beta x_\lambda  \gamma_5\gamma_\nu\gamma_\beta\gamma_\rho\gamma_{\omega}\gamma_\mu
\gamma_{\lambda} \right. 
\nonumber \\ 
\left. \pm s_2 y_\alpha x_\beta y_\lambda  \gamma_5\gamma_\nu\gamma_\beta\gamma_\rho\gamma_{\omega}\gamma_\mu
\gamma_{\lambda} 
\mp s_1 x_\alpha y_\beta x_\lambda  \gamma_5\gamma_\nu\gamma_{\omega}\gamma_\rho\gamma_{\beta}\gamma_\mu
\gamma_{\lambda} 
\pm s_1 x_\alpha y_\beta y_\lambda  \gamma_5\gamma_\nu\gamma_{\omega}\gamma_\rho\gamma_{\beta}\gamma_\mu
\gamma_{\lambda} \right) \nonumber \\ 
-i x_\alpha y_\beta x_\lambda \gamma_\nu\gamma_{\alpha}\gamma_\rho\gamma_{\beta}\gamma_\lambda 
+ i x_\alpha y_\beta y_\lambda \gamma_\nu\gamma_{\alpha}\gamma_\rho\gamma_{\beta}\gamma_\lambda \Big] \Big\rangle ~\epsilon^{a}_{\mu}(k)\epsilon^{b}_{\nu}(p)\epsilon_{\rho}(q), \nonumber 
\end{gather}
where $q_{f}$ is the ratio of quark electric charge to the electron charge.
The terms with the product of an even number of tensor 
$\hat{f}_{\mu\nu}$ in  $M^{(I)}$ and $M^{(II)}$ have opposite signs, while the signs of the terms with an odd number of 
$\hat{f}_{\mu\nu}$ coincide. In addition, the amplitudes  $M^{(I)}$ and $M^{(II)}$ differ by the sign of the phase factor: $e^{i\hat{f}_{\mu\nu}J_{\mu\nu}}$ for diagram $(I)$ and $e^{-i\hat{f}_{\mu\nu}J_{\mu\nu}}$ for diagram $(II)$. The sign of the phase factor is reflected in the result of averaging over the spatial orientation of the background field (for details see Ref.~\cite{Nedelko:2016vpj})
\begin{gather}
\Big\langle\prod_{j=1}^n f_{\alpha_j\beta_j}e^{\pm if_{\mu\nu}J_{\mu\nu}}\Big\rangle=\frac{(\pm 1)^n}{(2i)^n}\prod_{j=1}^n \frac{\partial}{\partial J_{\alpha_j\beta_j}}\frac{\sin\sqrt{2\left(J_{\mu\nu}J_{\mu\nu}\pm J_{\mu\nu}\widetilde{J}_{\mu\nu}\right)}}{\sqrt{2\left(J_{\mu\nu}J_{\mu\nu}\pm J_{\mu\nu}\widetilde{J}_{\mu\nu}\right)}},
\end{gather}
and
\begin{gather}
\Big\langle\prod_{j=1}^n f_{\alpha_j\beta_j}e^{-if_{\mu\nu}J_{\mu\nu}}\Big\rangle=(-1)^n \Big\langle\prod_{j=1}^n f_{\alpha_j\beta_j}e^{if_{\mu\nu}J_{\mu\nu}}\Big\rangle.
\end{gather}
Thus,  the terms in $M^{(I)}$ and $M^{(II)}$ with the product of an even number of the tensor $\hat{f}_{\mu\nu}$ cancel each other out  identically just as in the case of the ``usual" Furry theorem in QED, and the terms with the product of an odd number of the field strength tensor cancel each other upon averaging. The amplitude $M=M^{(I)}+M^{(II)}$ vanishes  in the confinement phase where averaging over random ensemble of almost everywhere homogeneous (anti-)self-dual vacuum gluon fields must be applied. The conversion of two gluons to a photon
does not occur in the presence of the random ensemble of confining vacuum fields.

\section{Chromomagnetic gluon fields and photon production in deconfined phase}
Within the mean-field approach to QCD vacuum (Refs.~\cite{Galilo:2011nh,Nedelko:2014sla}) and the lattice QCD studies (Refs.~\cite{DElia:2012ifm,Bali:2013esa}) it has been indicated that the strong magnetic field generated in relativistic heavy-ion collisions can play the role of a trigger for deconfinement transition to the phase characterized by the anisotropic chromomagnetic field. The chromomagnetic field $B$ exists as long as the deconfined phase persists, much longer than the initial pulse of the magnetic field $B_{\rm el}$. This is consistent with the indications that in the deconfining phase the scalar gluon condensate $\langle F^2\rangle$ remains nonzero above the critical temperature  while the mean absolute value of the topological charge  density (or, equivalently, the condensate $\langle (F\tilde F)^2\rangle$) turns to zero. A relevance of nonzero absolute value of topological charge density to confinement has been discussed in lattice QCD (e.g., see Refs.~\cite{Faber:1994fi,Moran:2008xq,Bornyakov:2013iva,Astrakhantsev:2019wnp,Lombardo:2020bvn}). 

The probable lifetime range $\Delta t$ of the magnetic field $B_{\rm el}$ in heavy-ion collisions depends on several factors: the total energy of the colliding ions $\sqrt{s_{NN}}$, the centrality class and the type of nuclei (Au+Au or Cu+Cu collisions). According Refs.~\cite{Shovkovy:2012zn,Ayala:2019jey} the lifetime range of the anisotropic magnetic field is $ \Delta t \le 1 $ fm for Au + Au collisions at $\sqrt{s_{NN}}=200$ GeV in the centrality class $0-40 \%$ and the maximum strength $B_{\rm el}$ is observed at the interval $ 0.1 \le \Delta t \le 0.2 $ fm. Bearing in mind that the chromomagnetic field $B$ exists much longer than the initial pulse of the magnetic field $B_{\rm el}$, it makes sense to  estimate the magnitude of the photon generation effect due to the gluon conversion in the background constant homogeneous chromomagnetic field, which models the limiting case of the maximum anisotropy in the system.

\begin{figure}[!h]
\center{\includegraphics[scale=0.5]{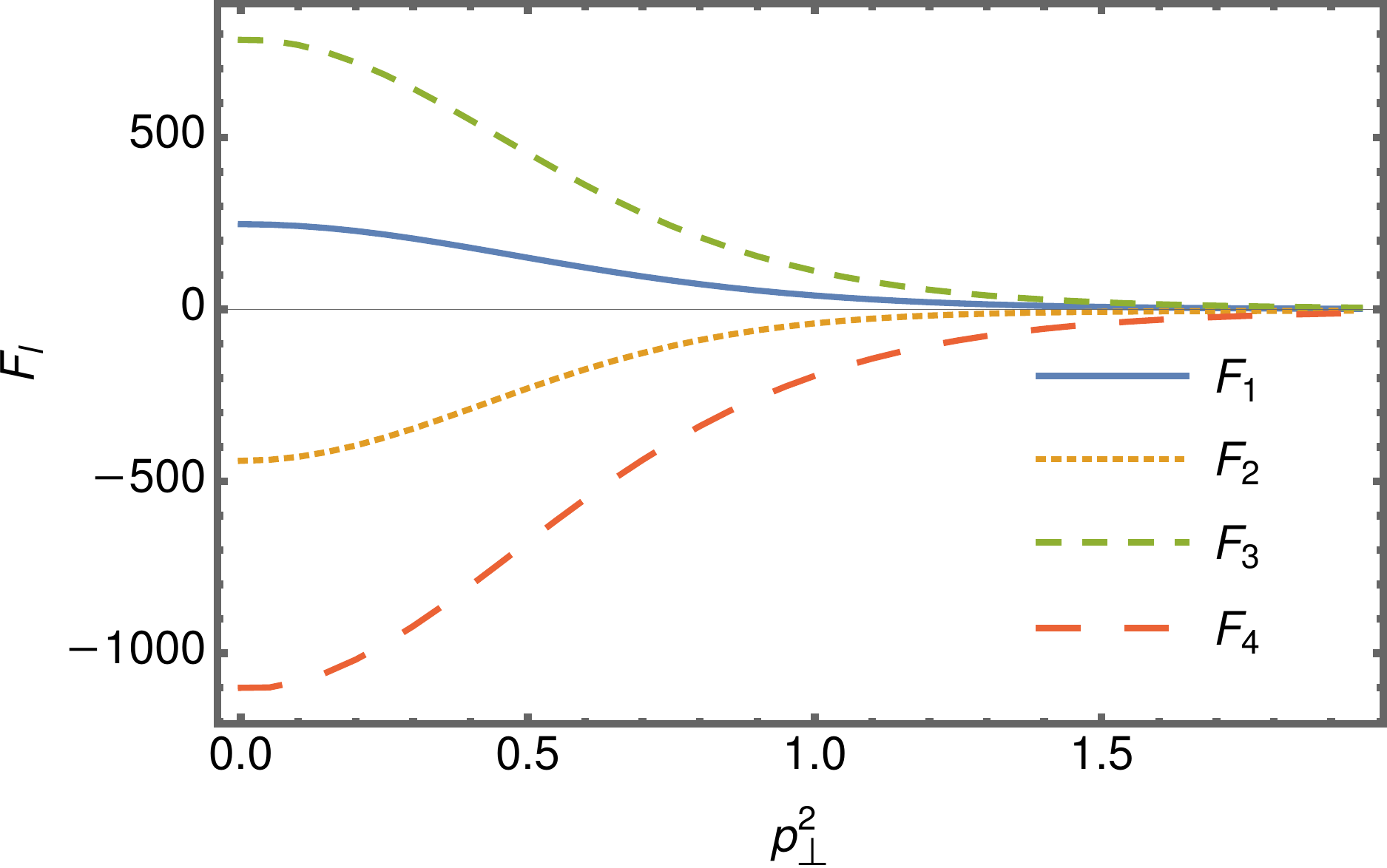}}
\caption{Some form factors $F_l$ in Eq.~\eqref{M_Fn} as a function of transverse gluon momenta $p^2_\perp=k^2_\perp$ for longitudinal  momenta $p^2_{||}=k^2_{||}=1$. Dimensionless notation $p^2=p^2/B$, $k^2=k^2/B$ is used, form factors $F_l$ are dimensionless. See Appendix for detailed form of $F_l$.}
\label{Fn_crhom}
\end{figure}

As it has already been mentioned, the emergent long-lived chromomagnetic field and initially generated magnetic field are expected to be parallel to each other (Refs.~\cite{Galilo:2011nh,Nedelko:2014sla,Bali:2013esa,DElia:2012ifm,Bonati:2018uwh}). This statement is based on the already well-tested observation that a strong magnetic field (or, more generally, an electromagnetic field with orthogonal magnetic and electrical components) introduces anisotropy into the dominant vacuum configurations (strong background fluctuations of gluon fields) of the gluon field, and thus serves as a  trigger for deconfinement, the so-called magnetic  catalysis for deconfinement. Namely, the vacuum configurations of the gluon field tend to repeat the configuration of the external electromagnetic field.  It is convenient to select the third spatial axis $x_3$ along the direction of the background (chromo)magnetic field:   
\begin{gather}
\label{chr-field}
\hat B_{\mu\nu}=\hat{n}B_{\mu\nu}=\hat{n}Bf_{\mu\nu}, \ f_{12}=-f_{21}=1,
\nonumber
\end{gather}
all other components of $f_{\mu\nu}$ are equal to zero. 
Respectively,  it is convenient to denote: 
\begin{gather}
\label{chr-spatial}
x_{\perp}=\big(x_{1},x_{2},0,0\big),~x_{||}=\left(0,0,x_{3},x_{4}\right). \nonumber
\end{gather}
The color charged quasi-particles  with ``masses'' $\mu_n$ defined by the Landau levels  can freely move along the chromomagnetic field and are confined in the transverse direction. Respectively, it is convenient to introduce notation for  longitudinal $p_{||}$ and transverse $p_{\perp}$ momenta  (in Euclidean space-time):
\begin{gather}
p_{\perp}=\left(p_{1},p_{2},0,0\right),~p_{||}=\left(0,0,p_{3},p_{4}\right). \nonumber 
\end{gather} 
The complete propagator of the quark field with mass $m_f$ in the presence of an external chromomagnetic field , accounting for contribution of all Landau levels, has the form
\begin{gather}
\label{chr-propagator}
S(x,y)=\exp\left\{-\frac{i}{2}x^{\mu}_{\perp}\hat B_{\mu\nu}
y^{\nu}_{\perp}\right\} H_f(x-y), \\
H_f(z) = \frac{B|\hat{n}|}{16\pi^{2}}\int\limits_0^{\infty}\frac{ds}{s}
\left[\coth(B|\hat{n}|s)-\sigma_{\rho\lambda}f_{\rho\lambda}\right]
\exp\left\{-m_f^{2}s-\frac{1}{4s}z_{||}^{2}-\frac{1}{8s}\left[B|\hat{n}|s\coth(B|\hat{n}|s)+1\right]z_{\perp}^{2}\right\}
\nonumber \\ 
\left\{m_f-\frac{i}{2s}\gamma_\mu z_{||}^{\mu}-\frac{1}{2}\gamma_{\mu}\hat B_{\mu\nu}z_{\perp}^{\nu}
-\frac{i}{4s}\left[B|\hat{n}|s\coth(B|\hat{n}|s)+1\right]\gamma_\mu z_\perp^{\mu}\right\}, \nonumber \\
\sigma_{\rho\lambda}=\frac{i}{2}\left[\gamma_\rho,\gamma_\lambda\right].\nonumber
\end{gather} 
The amplitudes corresponding to diagrams $(I)$ and $(II)$ in 
Fig.~\ref{th_loop} take the form
\begin{gather}
M^{(I)}=i(2\pi)^{4}\delta^{(4)}(p+k-q)g^{2}e\sum\limits_{f}q_f\int d^{4}xd^{4}y ~e^{-i(px+ky)-\frac{i}{2}\hat{n}By^{\mu}_{\perp}f_{\mu\nu}x^{\nu}_{\perp}} \nonumber \\
\label{Ma1chr}
\times \mathrm{Tr}\left[\gamma_{\nu}t^{b}H_f(x)\gamma_{\rho}H_f(-y)\gamma_{\mu}t^{a}H_f(y-x)\right]\epsilon^{a}_{\mu}(k)\epsilon^{b}_{\nu}(p)\epsilon_{\rho}(q), \\
M^{(II)}=i(2\pi)^{4}\delta^{(4)}(p+k-q)g^{2}e\sum\limits_{f}q_f\int d^{4}xd^{4}y e^{-i(px+ky)-\frac{i}{2}\hat{n}Bx^{\mu}_{\perp}f_{\mu\nu}y^{\nu}_{\perp}} \nonumber \\
\label{Mb1chr}
\times \text{Tr}\left[H_f(x-y)\gamma_{\mu}t^{a}H_f(y)\gamma_{\rho}H_f(-x)\gamma_{\nu}t^{b}\right]\epsilon^{a}_{\mu}(k)\epsilon^{b}_{\nu}(p)\epsilon_{\rho}(q). 
\end{gather} 
Further we consider the conversion of the gluons with their color orientation along the direction of the background field defined by the color vector $n^a=\delta^{a8}$. This particular case  corresponds to substitution  $t^a\epsilon_\mu^a\to t^8\epsilon_\mu^8$,  $t^b\epsilon_\nu^b\to t^8\epsilon_\nu^8$  in Eqs.~\eqref{Ma1chr} and \eqref{Mb1chr} ($t^8$ is defined in Eq.~\eqref{self-field}). These gluons do not interact with the background chromomagnetic field and  can be called ``neutral" with respect to the background field. Calculation of the Dirac trace and integration over variables $x_{||}$, $x_{\perp}$, $y_{||}$, $y_{\perp}$ leads to   
\begin{gather}
M=M^{(I)}+M^{(II)}=i(2\pi)^{4}\delta^{(4)}(p+k-q)g^{2}e \sum_l  \mathcal{F}^{l}_{\mu\nu\rho}(p,k)F_l(p,k)  \epsilon^{8}_{\mu}(k)\epsilon^{8}_{\nu}(p)\epsilon_{\rho}(q), 
\label{MFF}
\end{gather} 
where tensors $\mathcal{F}^{l}$ are composed of the $\delta_{\alpha\beta}$, momenta $p_\alpha$, $k_\alpha$ and the tensor $f_{\alpha\beta}$. Form factors $F_l$ have the structure 
\begin{gather}
F_{l}(p,k)= \sum_f q_f \mathrm{Tr}_{\hat{n}}\int^{\infty}_{0} ds_{1}ds_{2}ds_{3}\left[\psi^{(I)}_l\left(s_{1},s_{2},s_{3}|\hat{n},m_f\right) + 
\psi^{(II)}_l\left(s_{1},s_{2},s_{3}|\hat{n},m_f\right)\right] 
\label{M_Fn} \\ 
\times \exp\left\{ -p^{2}_{||}\phi_{1}(s_{1},s_{2},s_{3})-p_{||}k_{||}\phi_{2}(s_{1},s_{2},s_{3})-k^{2}_{||}\phi_{3}(s_{1},s_{2},s_{3})
\right. \nonumber \\ 
\left.
-p^{2}_{\perp}\phi_{4}(s_{1},s_{2},s_{3})-p_{\perp}k_{\perp}\phi_{5}(s_{1},s_{2},s_{3})-k^{2}_{\perp}\phi_{6}(s_{1},s_{2},s_{3})-m^{2}_f(s_{1}+s_{2}+s_{3})\right\}, \nonumber
\end{gather}
where  $\psi_l^{(I/II)}$ are the functions of $s_i$, and the functions $\phi_1,..,\phi_6$ in Eq.~\eqref{M_Fn} read
\begin{eqnarray}
&&\phi_1=\frac{t_1\left(t_2+t_3\right)}{t_1+t_2+t_3}, \ 
\phi_2=\frac{2t_1t_2}{t_1+t_2+t_3},  
\phi_3=\frac{t_2\left(t_1+t_3\right)}{t_1+t_2+t_3}, \  
 \nonumber \\
&&\phi_4= \frac{\xi_1\left(\xi_2+\xi_3\right)}{\xi_1+\xi_2+\xi_3+\xi_1\xi_2\xi_3}, \ \phi_5= \frac{2\xi_1\xi_2}{\xi_1+\xi_2+\xi_3+\xi_1\xi_2\xi_3}, ~ 
\phi_6= \frac{\xi_2\left(\xi_1+\xi_3\right)}{\xi_1+\xi_2+\xi_3+\xi_1\xi_2\xi_3}, \nonumber\\
&& t_j=B\left|\hat{n}\right|s_j, \ \ \xi_j=\frac{2t_j}{t_j\coth(t_j)+1}.
\label{xi}
\end{eqnarray}
Full expressions for amplitude $M$ in Eq.~\eqref{MFF} and form factors $F_{l}(p,k)$ in Eq.~\eqref{M_Fn} are given in the Appendix. Some form factors in Euclidean kinematics are shown in Fig.~\ref{Fn_crhom}. 

To calculate the on-shell amplitude squared $T=|M|^2$ one has to continue Eq.~\eqref{MFF} to Minkowsky kinematics of the gluon and photon momenta: 
\begin{gather}
\label{Wick-rot}
p^{2}_{||} \rightarrow - p^{2}_{||}, \ k^{2}_{||} \rightarrow - k^{2}_{||}, \ p_{||}k_{||} \rightarrow -p_{||}k_{||}. 
\end{gather} 
In Minkowski space-time, on-shell conditions for gluons and photon $p^2=0$, $(p+k)^2=0$ and $k^2=0$ impose the following relations  
\begin{gather}
\label{mass-shell}
p^{2}_{||} = p^{2}_{\perp}, \  k^{2}_{||} = k^{2}_{\perp}, \ p_{||}k_{||} = p_{\perp}k_{\perp} 
\end{gather}
and the exponential phase factor $ip^\mu_{\perp}f_{\mu\nu}k^\nu_{\perp}$  vanishes since vectors $p_{\perp}$ and $k_{\perp}$ are parallel to each other.

The probability of photon production is given by the squared amplitude  averaged over the initial gluon polarization states and summed over the final polarizations of the photon 
\begin{gather}
\overline T(p,k,q)=\Delta v\Delta\tau(2\pi)^4\delta^4(p+k-q)~T(p,k)~, \nonumber
\end{gather}
here $\Delta v\Delta\tau$ - is a space-time volume, 
\begin{eqnarray}
T(p,k)&=&\frac{2\alpha\alpha^{2}_{s}}{\pi}\int_{0}^{\infty}
ds_{1}ds_{2}ds_{3}dr_{1}dr_{2}dr_{3} \ F(s_{1},s_{2},s_{3},r_{1},r_{2},r_{3}|p,k) \nonumber \\
&\times&\exp \left\{ p_{\perp}^{2}\Phi_{1}(s_{1},s_{2},s_{3},r_1,r_2,r_3) + p_{\perp}k_{\perp}\Phi_2(s_{1},s_{2},s_{3},r_1,r_2,r_3) + \right. \nonumber \\ 
&&\left.
k_{\perp}^{2}\Phi_3(s_{1},s_{2},s_{3},r_1,r_2,r_3) -m^2_f(s_1+s_2+s_3+r_1+r_2+r_3) \right\}, 
\label{Tl}
\end{eqnarray}
where $\alpha $ and $\alpha_{s}$ --  electromagnetic and strong coupling constants, and 
\begin{gather} 
\Phi_1=\phi_{1}(s_{1},s_{2},s_{3})+\phi_{1}(r_{1},r_{2},r_{3})-\phi_{4}(s_{1},s_{2},s_{3})-\phi_{4}(r_{1},r_{2},r_{3}), \nonumber \\ 
\Phi_2=\phi_{2}(s_{1},s_{2},s_{3})+\phi_{2}(r_{1},r_{2},r_{3})-\phi_{5}(s_{1},s_{2},s_{3})-\phi_{5}(r_{1},r_{2},r_{3}), \nonumber \\
\Phi_3=\phi_{3}(s_{1},s_{2},s_{3})+\phi_{3}(r_{1},r_{2},r_{3})-\phi_{6}(s_{1},s_{2},s_{3})-\phi_{6}(r_{1},r_{2},r_{3}). 
\label{phiL} 
\end{gather}
Pre-exponential factor $F$ is a polynomial in $p_\perp^2$, $k_\perp^2$, $k_\perp p_\perp$ with  coefficients being the rational functions of proper times $(s_j,r_j)$ and their combinations $(\xi_j(s_j),\xi_j(r_j))$ given in Eq.~\eqref{xi}. 

Since the functions $\Phi_j$ are positive in the whole region of  integration and grow linearly for $s_j\to\infty$, the proper time integrals in Eq.~\eqref{Tl} converge only for the limited range of momenta $p_\perp$ and $k_\perp$. For instance, for the case $k_\perp^2=p_\perp^2$ the integral  converges if  
\begin{eqnarray}
p_\perp^2<\frac{3}{2}m_f^2,
\label{conditionconv}
\end{eqnarray}
as it can be seen from Eqs.~\eqref{xi} and \eqref{phiL}. 
 
A straightforward numerical calculation of the function $T(p,k)$ in Eq.~\eqref{Tl} by means of  analytical continuation is a long standing technical problem. A profound consideration of the almost identical task related to studying  the analytical  properties of the amplitude corresponding to the triangle diagram for  photon splitting in the external electromagnetic field can be found in Ref.~\cite{Papanyan:1973xa} (see also Refs.~ \cite{Ritus1970,Ritus:1972ky,Papanyan:1971cv,Adler:1970gg} for simpler case of two-point vacuum polarization in external magnetic field). Thus, though a general prescription for analytical continuation of Eq.~\eqref{Tl} to the arbitrary values of momenta in the complex plane has been formulated a long time ago by Papanyan and Ritus, its practical application is rather complicated and has to be elaborated yet. So far we shall use Eq.~\eqref{Tl} for computation in the region of its applicability, limited by the condition in Eq.~\eqref{conditionconv}, just for comparison with the approximate result based on the decomposition of the quark propagators accounting for the lowest  Landau levels. 

The main purpose of the present paper is to demonstrate, in principle, the effect of a long-range chromomagnetic field on the generation of photons in the deconfinement phase during relativistic collisions of heavy ions, and for this it is sufficient  to consider the limit of the strong field and small quark masses. In the massless quark limit  the process of two gluon conversion to a photon in the presence of long-range magnetic field was studied in Ref.~\cite{Ayala:2019jey} with the following result for the amplitude squared (the case of a single flavour) accounting for the lowest Landau levels (LLL) and first excited Landau level (1LL) for the quark propagator:   
\begin{gather}
\label{T-mex}
T(p,k)=\frac{2\alpha\alpha^{2}_{s}}{\pi}~q^2_f\left(2p_{\perp}^{2}+ k_{\perp}^{2} + p_{\perp}k_{\perp}\right)\exp\left\{-\frac{1}{|q_fB_{\rm el}|} \big(p_{\perp}^{2}+k_{\perp}^{2}+p_{\perp}k_{\perp}\big) \right\},
\end{gather}
where $B_{el }$ - is the strength of magnetic field.

\begin{figure}[!h]
\center{\includegraphics[scale=0.5]{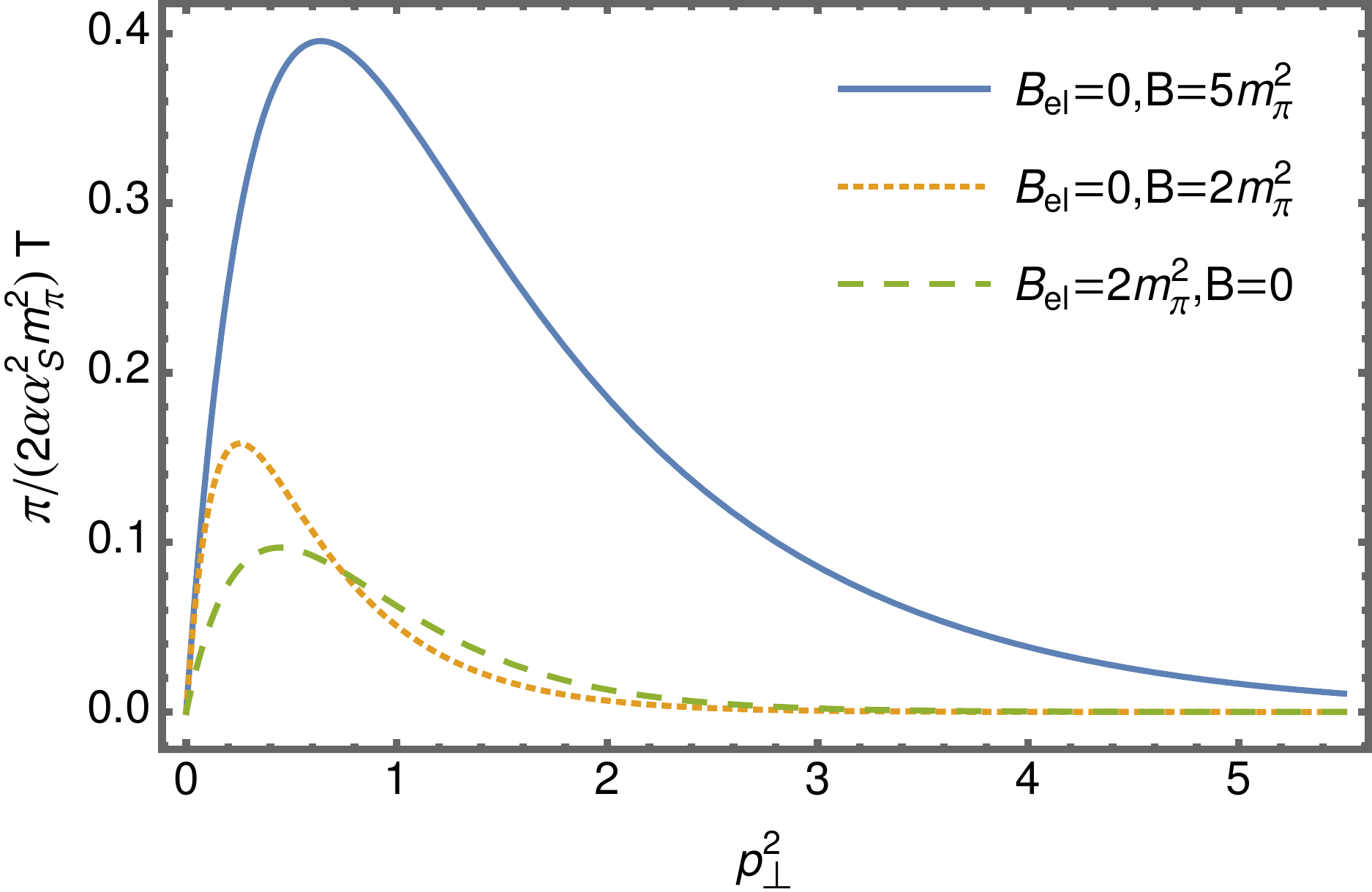}}
\caption{Dependence of $T(p,k)$ given by Eq.~\eqref{T-mex-chrom} in regime $k^2_\perp=p^2_\perp$. The dashed line corresponds to the purely magnetic field $B_{\rm el}$, dotted and solid lines represent the case of pure chromomagnetic  field $B$ with different values of strength. The contribution of massless  $u$-quark is taken into account and the mass of the pion $m_{\pi}$ is chosen as the scale. Dimensionless notation $p^2_\perp=p^2_\perp/B$ is used. } 
\label{fig3}
\end{figure} 

Equation \eqref{T-mex} can be easily generalized to the case of the presence of both magnetic and chromomagnetic fields  by the replacement  $|q_fB_{\rm el}|\to |q_fB_{\rm el}+\hat{n}B|$ with the result
\begin{gather}
\label{T-mex-chrom}
T=\frac{2\alpha\alpha_{s}^2}{N_c\pi} q^2_f ~ 
\text{Tr}_{\hat{n}} \left(2p_{\perp}^{2}+ k_{\perp}^{2} + p_{\perp}k_{\perp}\right)\exp\left\{-\frac{p_{\perp}^{2}+k_{\perp}^{2}+p_{\perp}k_{\perp}}{|q_fB_{\rm el}+\hat{n}B|}  \right\},
\end{gather}
where the angle between magnetic and chromomagnetic fields is assumed to be zero, which corresponds to a minimum of the one-loop quark contribution to the free energy density (Ref.~\cite{Galilo:2011nh}).

\begin{figure}[!h]
\center{\includegraphics[scale=0.5]{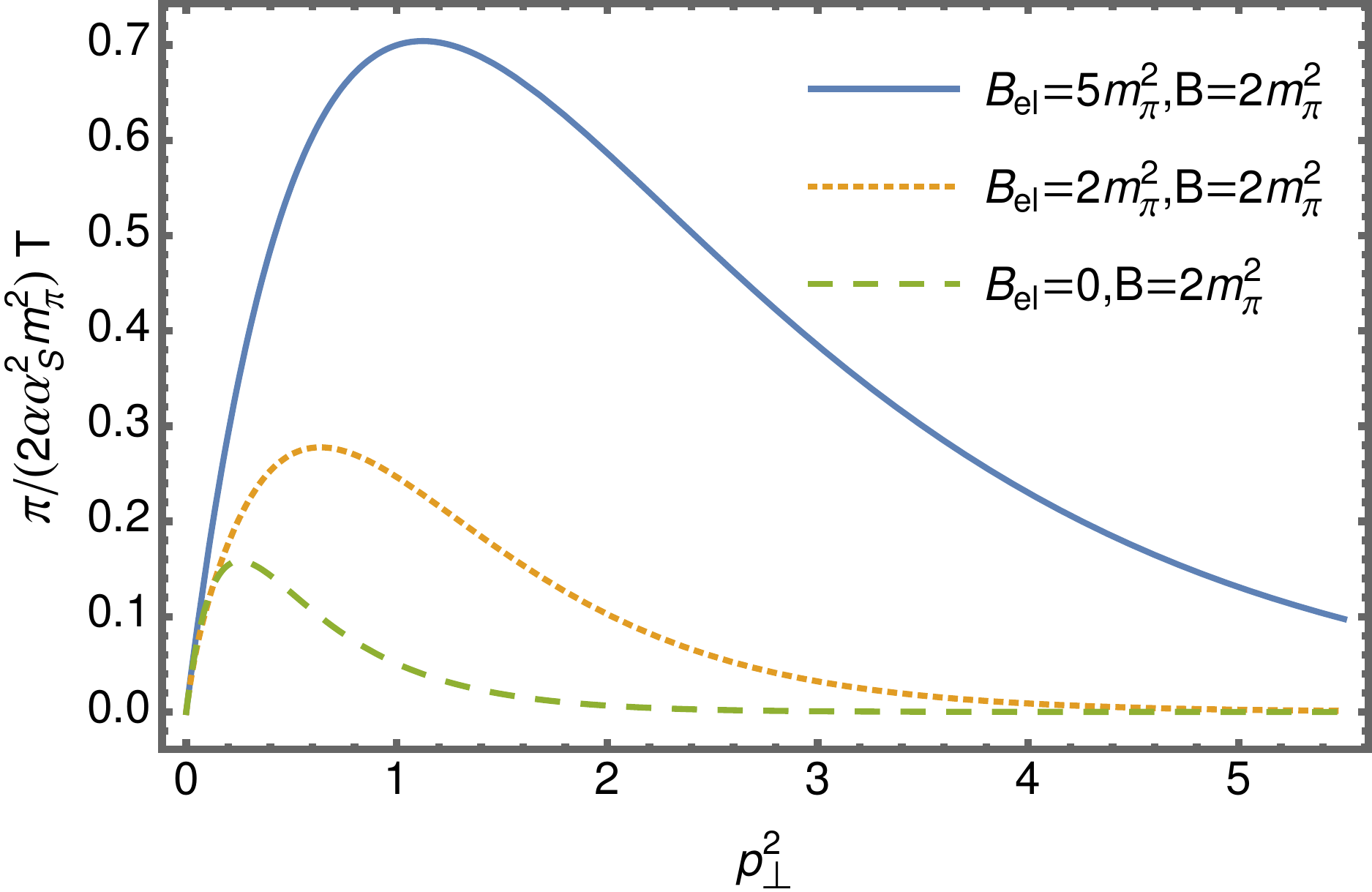}}
\caption{Dependence of $T(p,k)$ given by Eq.~\eqref{T-mex-chrom} on the strength of magnetic field $B_{el}$ in regime $p^2_\perp=k^2_\perp$. The dashed curve corresponds to  the presence of chromomagnetic field $B=2m^2_{\pi}$ alone,  the dotted and solid curves represent the effect of  addition of  a magnetic field.  Dimensionless notation $p_\perp=p^2_\perp/B$ is used. } 
\label{fig4}
\end{figure}

As it is seen from the comparison of dotted and dashed lines  in Fig.~\ref{fig3}, the chromomagnetic field enhances the photon production amplitude at small $p_\perp$ in comparison with the effect of pure magnetic field with the same strength as the chromomagnetic one.  One may expect that chromomagnetic field should be much stronger than the magnetic one,  since the strength of the chromomagnetic field squared is of order of the value of the scalar gluon condensate. If so then the effect of chromomagnetic field can be strong over a wide range of gluon momenta, see solid line in Fig.~\ref{fig3}. Since the magnetic field strength $B_{\rm el}$ decreases soon after the heavy-ion collisions but the chromomagnetic field strength $B$ is expected to be constant in the deconfined phase, then at a certain moment the production of photons will be determined only by the presence of the chromomagnetic field, see Fig.~\ref{fig4}. 

Using relations between the momenta $p,k,q$ and the energies $\omega_p,\omega_k,\omega_q$
\begin{gather}
p^\mu=(\omega_p / \omega_q)q^\mu , \nonumber \\
k^\mu=(\omega_k / \omega_q)q^\mu, 
\end{gather}
one can rewrite Eq.~\eqref{T-mex-chrom} in the form
\begin{gather}
\label{T-chrom-mex-omega}
T(\omega_p,\omega_k)=\frac{2\alpha\alpha^{2}_{s}q^2_\perp}{N_c \pi\omega^2_q}q^2_f{\rm Tr}_{\hat n}\left(2\omega^2_p + \omega^2_k + \omega_p\omega_k\right)\exp\left\{-\frac{\left(\omega^2_p + \omega^2_k + \omega_p\omega_k\right)q^2_\perp}{|q_fB_{\rm el}+\hat n B|\omega^2_q}  \right\}.
\end{gather}

The invariant photon momentum distribution is thus given by 
\begin{gather}
\label{ivn-mom-distr}
\omega_q\frac{dN}{d^3q}=\frac{\Delta v\Delta\tau}{2(2\pi)^3} \int \frac{d^3p}{(2\pi)^3 2\omega_p} \int \frac{d^3k}{(2\pi)^3 2\omega_k} n(\omega_p)n(\omega_k) \delta^4(q-k-p) ~T(\omega_p,\omega_k), 
\end{gather}
where $n(\omega)$ represents the distribution of gluons. Following the argumentation of Ref.~ \cite{Ayala:2019jey} we shall use the distribution
\begin{gather}
\label{B-E-d}
n(\omega)=\frac{\eta}{e^{\omega/\Lambda_s}-1},
\end{gather}
where $\eta$ represents the high gluon occupation factor.
The factor $\Delta v\Delta\tau$ comes from squaring the delta function for energy-momentum conservation in amplitude. This factor represents the space-time volume where the reaction takes place and consists of the
product of the spatial volume of the nuclear overlap region $\Delta v(t)$ at time $t$ and the time interval $\Delta\tau$ where the magnetic and chromomagnetic fields can be taken as having a constant strength.

Finally, the invariant photon momentum distribution in the presence of chromomagnetic and magnetic fields can be represented in the form 
\begin{gather}
\frac{1}{2\pi\omega_q} \frac{dN}{d\omega_q}= \nu\Delta\tau \frac{\alpha\alpha^2_s \pi}{2N_c(2\pi)^6 \omega_q} q^2_f ~
\text{Tr}_{\hat{n}} \int_0^{\omega_q}d\omega_p \left(2\omega^2_p + \omega^2_q -\omega_p\omega_q\right) ~ e^{-g^{B}_f(\omega_p,\omega_q)} \nonumber \\
\label{dis-Bes-chrom-magn} 
\left[I_0\left(g^{B}_f(\omega_p,\omega_q)\right) - I_1\left(g^{B}_f(\omega_p,\omega_q)\right)\right] \left(n(\omega_p)n(|\omega_q-\omega_p|)\right) ,  
\end{gather}
where
\begin{equation}
g^{B}_f(\omega_p,\omega_q)=\frac{\omega^2_p+\omega^2_q-\omega_p\omega_q}{2|\hat{n}B+q_fB_{\rm el}|} \nonumber. 
\end{equation}
and $I_0\left(g_f(\omega_p,\omega_q)\right),~I_1\left(g_f(\omega_p,\omega_q)\right)$ - are the modified Bessel function of the first kind.

\begin{figure}[!h]
\center{\includegraphics[scale=0.5]{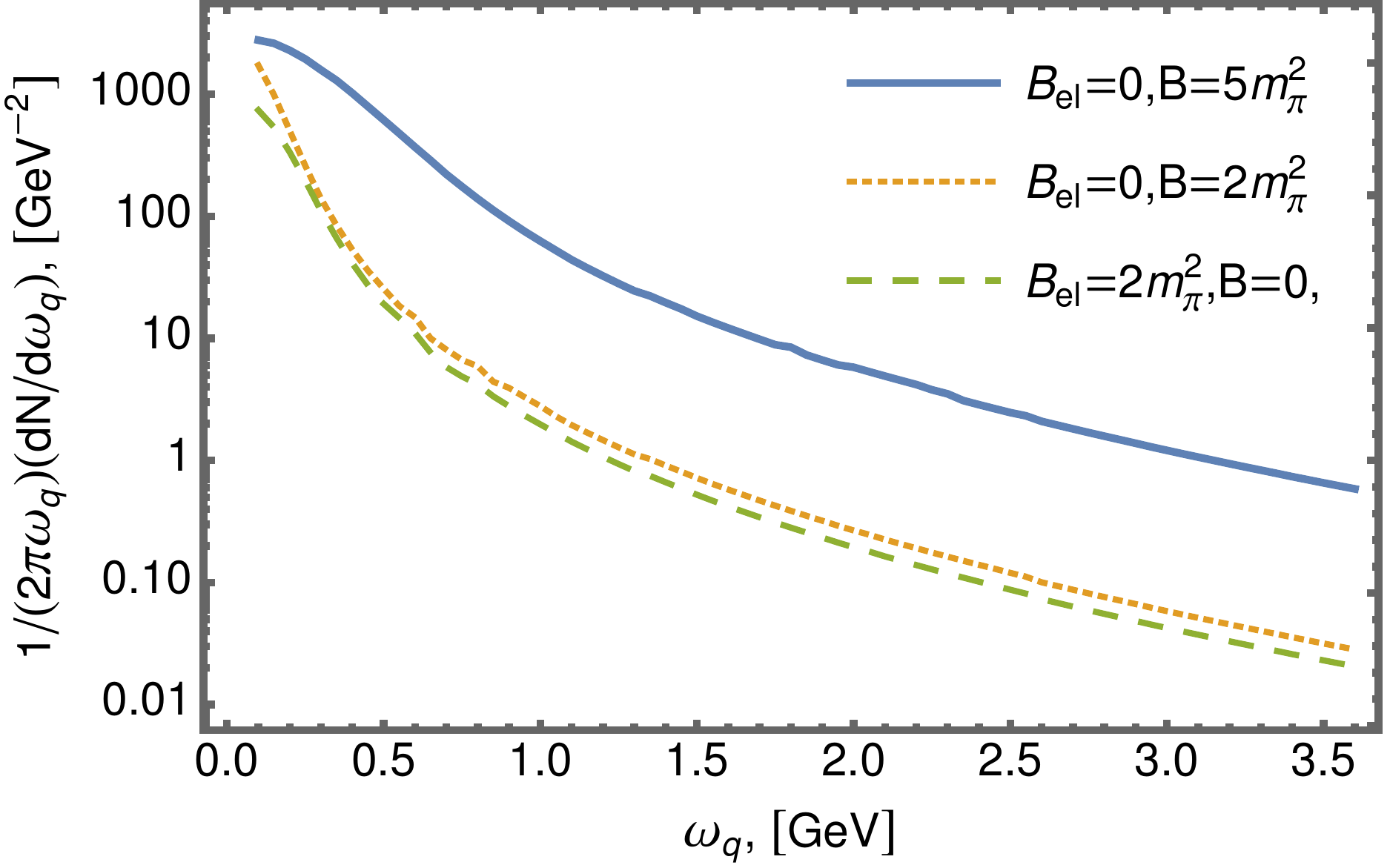}}
\caption{Differential energy distribution in Eq.~\eqref{dis-Bes-chrom-magn} of the generated photons for a pure magnetic field $B_{\rm el}$ (dashed line)  and pure chromomagnetic field $B$ (the dotted and solid curves). The pion mass $m_{\pi}=0.135$ GeV is chosen as the scale. The factor $\Delta v\Delta\tau\alpha\alpha^2_s/(2N_c(2\pi)^6)=1 \text{Gev}^{-4}$.} 
\label{fig5}
\end{figure}

A comparison of the differential energy distribution  of generated photons in the background magnetic field $B_{\rm el}$ and  chromomagnetic field $B$ is shown in Fig.~\ref{fig5}. Note that the integral in Eq.~\eqref{dis-Bes-chrom-magn} is regularized at the lower limit using the infrared scale $\Lambda_{IR}=0.05$ GeV which corresponds to thermal gluon distribution as has been defined in Ref.~\cite{McLerran:2014hza}.

\begin{figure}[!h]
\center{\includegraphics[scale=0.5]{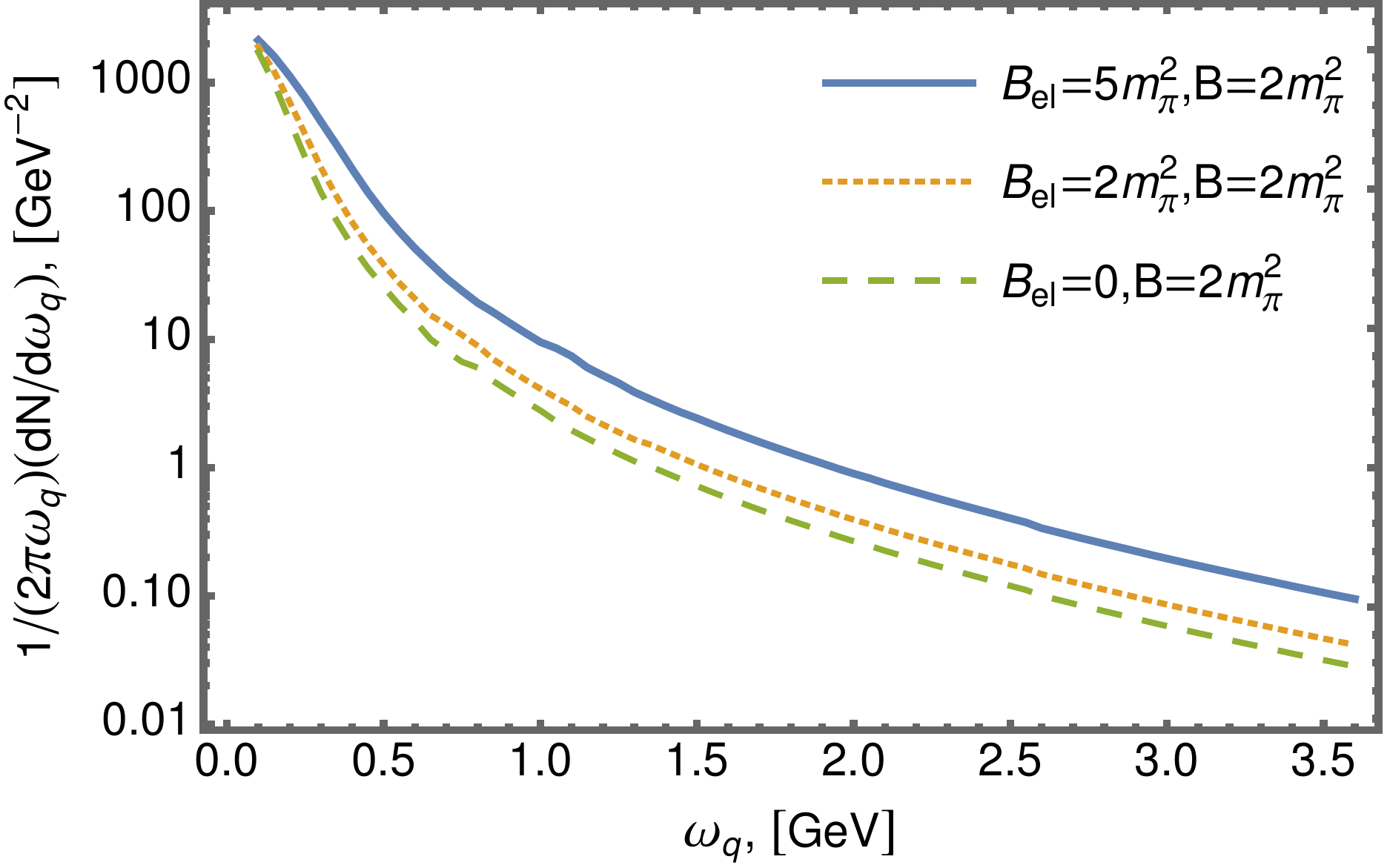}}
\caption{Differential energy distribution in Eq.~\eqref{dis-Bes-chrom-magn} of the generated photons. The dashed curve corresponds to chromomagnetic field $B=2m^2_{\pi}$ alone, and the dotted and solid curves correspond to addition of magnetic field. } 
\label{fig6}
\end{figure}

The response of the invariant photon momentum distribution in Eq.~\eqref{dis-Bes-chrom-magn} to a change in the magnetic field strength is shown in Fig.~\ref{fig6},  where magnetic field strength $B_{\rm el}$ varies, but the chromomagnetic field strength $B$ stay unchanged. It is clearly seen that a decrease in the magnetic field strength leads to a decrease in the level of the photon production signal and is further determined  by the strength of the chromomagnetic field. Thus the photon production occurs as long as the deconfined phase exists irrespective to the disappearance  of the initial  magnetic field. As a matter of fact,  the main role of magnetic field is to trigger the anisotropy of chromomagnetic field, which than stays as long as the deconfined phase exits.

\begin{figure}[!h]
\center{\includegraphics[scale=0.5]{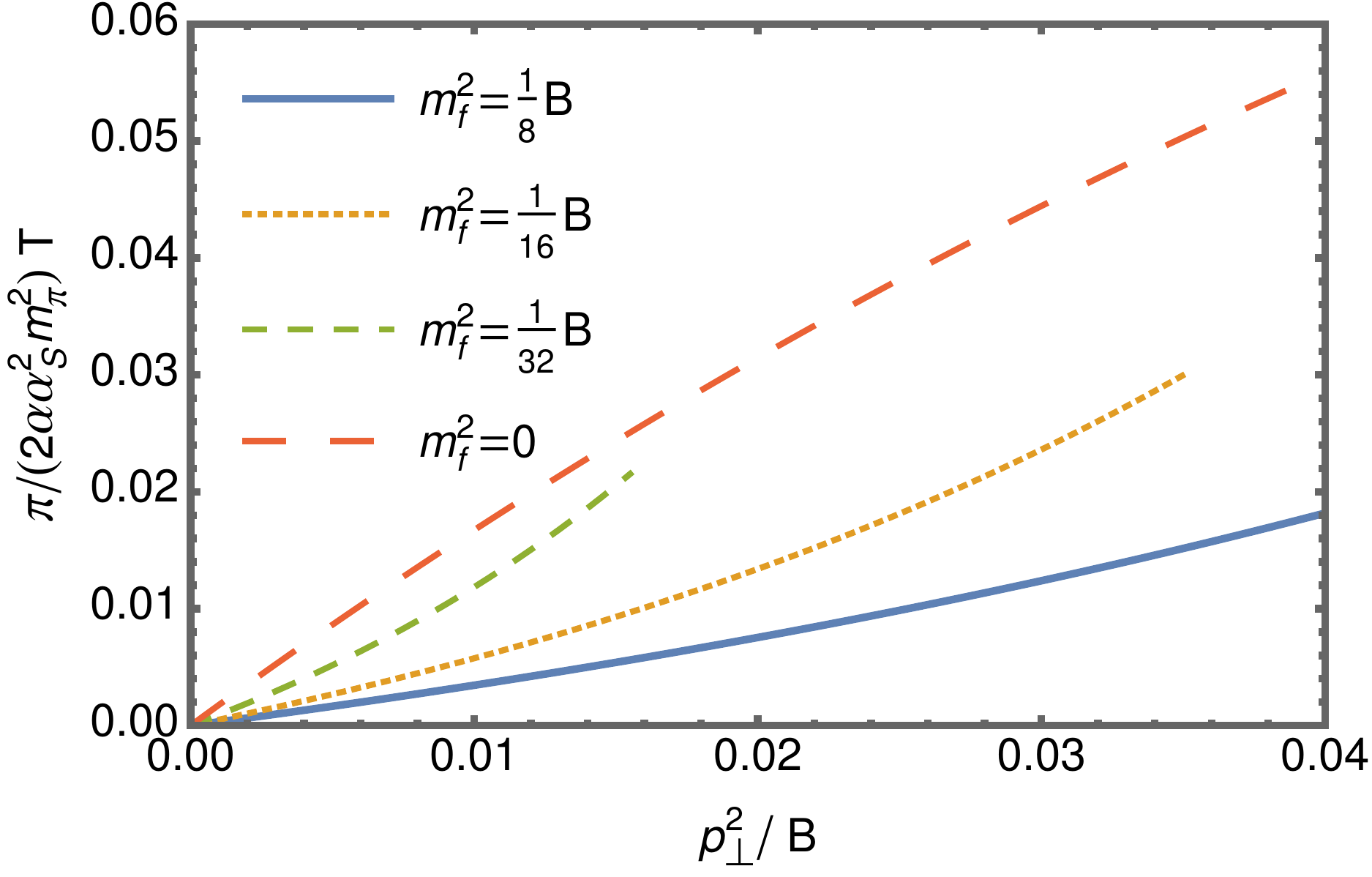}}
\caption{The comparison of the amplitude square obtained by the Landau level decomposition for a massless quark, Eq.~\eqref{T-mex-chrom}, with the amplitude square, Eq.~\eqref{Tl}, taking into account all Landau levels for different quark masses $m_f$ at low gluon momenta in the regime $k^2_\perp=p^2_\perp<3m_f^2/2$. The chromomagnetic field strength $B=4m_\pi^2$ and magnetic field $B_{el}=0$. Dimensionless notation $p^2_\perp=p^2_\perp/B$ is used.} 
\label{fig7}
\end{figure}

\newpage

\section{Discussion and outlook}

The estimates presented in this paper have illustrated  at the qualitative level the plausibility of a peculiar mechanism of photon generation in the quark-gluon plasma. In heavy-ion collisions the conditions of  Furry theorem are not satisfied due to the presence of strong anisotropic electromagnetic and chromomagnetic fields. Chromomagnetic field is likely to persist for a long time in comparison with a pure magnetic field (for details see Refs.~\cite{Galilo:2011nh,Nedelko:2014sla}). This mechanism is a kind of extension of the scenario  intensively  discussed in Refs.~\cite{Ayala:2017vex,Ayala:2019jey}. Indeed, photon production in the process $ gg \rightarrow \gamma $ may serve as a signal of the transition from the confinement to the deconfinement phase. 

We have not attempted a comparison with available experimental data just for the reason that quantitative level of consideration requires, at least, taking into account nonzero quark masses, especially for the strange quark. For strange quark a  contribution of all Landau levels has to be accounted since the decomposition over Landau levels may become unreliable if the value of a quark mass squared is of order of the strength of the background gauge field.  An integral representation for such a complete expression for the amplitude of the process $ gg \rightarrow \gamma $  was derived and is given in Eq.~\eqref{MFF} and Appendix. 

In order to estimate the deviation of the ``complete'' result from the  the approximation based  on taking into account two lowest  Landau levels with massless quarks, it makes sense to compute the amplitude in Eq.~\eqref{MFF} for the interval $k^2_\perp=p^2_\perp<3m_f^2/2$, where this representation is applicable,  and compare it with the approximate result. Such a comparison is given in Fig.~\ref{fig7}, which demonstrates an importance of contributions coming from the quarks with a mass of order of the current mass of the strange quark (solid line), as well as certain deviation of the light quark contribution (short dashed line). 

As it has already been mentioned, the obtained proper time integral representation for the probability of photon production in Eq.~\eqref{Tl} is not appropriate for computation of photon distribution in the whole physically interesting interval of photon momentum. Apparently, this problem is caused by using the complete quark propagator, i.e. accounting for contributions from all Landau levels, since taking into account only the lowest Landau levels of the quark does not lead to such difficulties (Refs.~\cite{Ayala:2017vex,Ayala:2019jey}). We note several possible ways to overcome the limitation fixed in Eq.~\eqref{conditionconv}, namely: more complex form of an analytical continuation into Minkowski space for Eq.~\eqref{MFF}, the transition to other integration variables in Eq.~\eqref{Tl} or the rotation of the integration contour in the complex plane for Eq.~\eqref{Tl}. The indicated methods for the transformation of Equation~\eqref{Tl} for arbitrary gluon momenta is under consideration and will be presented elsewhere. 

Of course, the lifetime and strength of the magnetic and chromomagnetic fields are very important for the studied photon generation process. The probable lifetime range of the magnetic field $B_{\rm el}$ in heavy-ion collisions depends on following drivers: the total energy of the colliding ions $\sqrt{s_{NN}}$, the centrality class and the type of nuclei (Au+Au or Cu+Cu collisions). According to Refs.~\cite{Shovkovy:2012zn,Ayala:2019jey} the lifetime range of the magnetic field is $ \Delta t \le 1 $ fm for Au + Au collisions at $\sqrt{s_{NN}}=200$ GeV in the centrality class $0-40 \%$. The maximum strength of the magnetic field $B_{\rm el}$ is observed at the range $ 0.1 \le \Delta t \le 0.2 $ fm and rapidly decreases. To estimate the lifetime of the chromomagnetic field $B$ one has to assume, that  the chromomagnetic field exists as long as the deconfinement phase persists. The reasoning for this assumption comes from the observation that the scalar gluon condensate seems to remain nonzero above the critical temperature while the mean absolute value of the topological charge density vanishes. The indications for a possible connection of the vanishing topological charge density with the confinement-deconfinement phase transition were obtained in the lattice QCD (Refs.~\cite{Astrakhantsev:2019wnp,Lombardo:2020bvn}). Within the mean-field picture of the conﬁning domain wall network one may speculate that during heavy-ion collisions a strong flash of the magnetic field  produces a thick domain wall junction in the conﬁning gluon background exactly in the region where collision occurs (see Ref.~\cite{Nedelko:2014sla} for details). The chromomagnetic field, which ``remembers'' the initial direction of the magnetic field, dominates in the region until the restoration of the confinement phase. 

\acknowledgments

We are grateful to Vladimir Voronin for numerous useful discussions and valuable comments.

\newpage

\section{Appendix}

The amplitude in Eq.~\eqref{MFF} consists of $32$ terms. Below an explicit form of the tensor structures and corresponding form factors are listed. 

Set of tensors $\mathcal{F}^{l}_{\mu\nu\rho}(p,k)$  includes
\begin{multline}
\mathcal{F}^{1}_{\mu\nu\rho}(p,k)=f_{\alpha \mu}f_{\beta \nu}f_{\lambda \rho}p_{\perp}^{\alpha}p_{\perp}^{\beta}p_{\perp}^{\lambda},~
\mathcal{F}^{2}_{\mu\nu\rho}(p,k)=f_{\alpha \mu}f_{\beta \nu}f_{\lambda \rho}p_{\perp}^{\alpha}p_{\perp}^{\beta}k_{\perp}^{\lambda},~ 
\mathcal{F}^{3}_{\mu\nu\rho}(p,k)=f_{\alpha \mu}f_{\beta \rho}f_{\lambda \nu}p_{\perp}^{\alpha}p_{\perp}^{\beta}k_{\perp}^{\lambda}  
\nonumber \\
\mathcal{F}^{4}_{\mu\nu\rho}(p,k)=f_{\alpha \mu}f_{\beta \nu}f_{\lambda \rho}p_{\perp}^{\alpha}k_{\perp}^{\beta}k_{\perp}^{\lambda},~  
\mathcal{F}^{5}_{\mu\nu\rho}(p,k)=f_{\alpha \nu}f_{\beta \rho}f_{\lambda \mu}p_{\perp}^{\alpha}p_{\perp}^{\beta}k_{\perp}^{\lambda},~  
\mathcal{F}^{6}_{\mu\nu\rho}(p,k)=f_{\alpha \nu}f_{\beta \mu}f_{\lambda \rho}p_{\perp}^{\alpha}k_{\perp}^{\beta}k_{\perp}^{\lambda},
\nonumber \\
\mathcal{F}^{7}_{\mu\nu\rho}(p,k)=f_{\alpha \rho}f_{\beta \mu}f_{\lambda \nu}p_{\perp}^{\alpha}k_{\perp}^{\beta}k_{\perp}^{\lambda},~ 
\mathcal{F}^{8}_{\mu\nu\rho}(p,k)=f_{\alpha \mu}f_{\beta \nu}f_{\lambda \rho}k_{\perp}^{\alpha}k_{\perp}^{\beta}k_{\perp}^{\lambda}, 
\nonumber \\
\mathcal{F}^{9}_{\mu\nu\rho}(p,k)=f_{\alpha \mu}p_{\perp}^{\alpha}\delta_{\nu\rho}p_{\perp}^2,~ 
\mathcal{F}^{10}_{\mu\nu\rho}(p,k)=f_{\alpha \mu}p_{\perp}^{\alpha}\delta_{\nu\rho}p_{\perp}k_{\perp}, ~ 
\mathcal{F}^{11}_{\mu\nu\rho}(p,k)=f_{\alpha \mu}p_{\perp}^{\alpha}\delta_{\nu\rho}k_{\perp}^2, 
\nonumber \\
\mathcal{F}^{12}_{\mu\nu\rho}(p,k)=f_{\alpha \mu}p_{\perp}^{\alpha}\delta^{||}_{\nu\rho},~    
\nonumber \\ 
\mathcal{F}^{13}_{\mu\nu\rho}(p,k)=f_{\alpha \nu}p_{\perp}^{\alpha}\delta_{\mu\rho}p_{\perp}^2, ~ 
\mathcal{F}^{14}_{\mu\nu\rho}(p,k)=f_{\alpha \nu}p_{\perp}^{\alpha}\delta_{\mu\rho}p_{\perp}k_{\perp},~ 
\mathcal{F}^{15}_{\mu\nu\rho}(p,k)=f_{\alpha \nu}p_{\perp}^{\alpha}\delta_{\mu\rho}k_{\perp}^2, 
\nonumber \\
\mathcal{F}^{16}_{\mu\nu\rho}(p,k)=f_{\alpha \nu}p_{\perp}^{\alpha}\delta^{||}_{\mu\rho},~
\nonumber \\ 
\mathcal{F}^{17}_{\mu\nu\rho}(p,k)=f_{\alpha \rho}p_{\perp}^{\alpha}\delta_{\mu\nu}p_{\perp}^2,~  
\mathcal{F}^{18}_{\mu\nu\rho}(p,k)=f_{\alpha \rho}p_{\perp}^{\alpha}\delta_{\mu\nu}p_{\perp}k_{\perp},~ 
\mathcal{F}^{19}_{\mu\nu\rho}(p,k)=f_{\alpha \rho}p_{\perp}^{\alpha}\delta_{\mu\nu}k_{\perp}^2, 
\nonumber \\
\mathcal{F}^{20}_{\mu\nu\rho}(p,k)=f_{\alpha \rho}p_{\perp}^{\alpha}\delta^{||}_{\mu\nu},~ 
\nonumber \\ 
\mathcal{F}^{21}_{\mu\nu\rho}(p,k)=f_{\alpha \mu}k_{\perp}^{\alpha}\delta_{\nu\rho}p_{\perp}^2,~ 
\mathcal{F}^{22}_{\mu\nu\rho}(p,k)=f_{\alpha \mu}k_{\perp}^{\alpha}\delta_{\nu\rho}p_{\perp}k_{\perp},~  
\mathcal{F}^{23}_{\mu\nu\rho}(p,k)=f_{\alpha \mu}k_{\perp}^{\alpha}\delta_{\nu\rho}k_{\perp}^2 , 
\nonumber \\
\mathcal{F}^{24}_{\mu\nu\rho}(p,k)=f_{\alpha \mu}k_{\perp}^{\alpha}\delta^{||}_{\nu\rho},~  
\nonumber \\ 
\mathcal{F}^{25}_{\mu\nu\rho}(p,k)=f_{\alpha \nu}k_{\perp}^{\alpha}\delta_{\mu\rho}p_{\perp}^2,~ 
\mathcal{F}^{26}_{\mu\nu\rho}(p,k)=f_{\alpha \nu}k_{\perp}^{\alpha}\delta_{\mu\rho}p_{\perp}k_{\perp},~ 
\mathcal{F}^{27}_{\mu\nu\rho}(p,k)=f_{\alpha \nu}k_{\perp}^{\alpha}\delta_{\mu\rho}k_{\perp}^2   
\nonumber \\
\mathcal{F}^{28}_{\mu\nu\rho}(p,k)=f_{\alpha \nu}k_{\perp}^{\alpha}\delta^{||}_{\mu\rho},~ 
\nonumber \\ 
\mathcal{F}^{29}_{\mu\nu\rho}(p,k)=f_{\alpha \rho}k_{\perp}^{\alpha}\delta_{\mu\nu}p_{\perp}^2,~ 
\mathcal{F}^{30}_{\mu\nu\rho}(p,k)=f_{\alpha \rho}k_{\perp}^{\alpha}\delta_{\mu\nu}p_{\perp}k_{\perp},~ 
\mathcal{F}^{31}_{\mu\nu\rho}(p,k)=f_{\alpha \rho}k_{\perp}^{\alpha}\delta_{\mu\nu}k_{\perp}^2,  
\nonumber \\
\mathcal{F}^{32}_{\mu\nu\rho}(p,k)=f_{\alpha \rho}k_{\perp}^{\alpha}\delta^{||}_{\mu\nu}, \nonumber \\ 
\end{multline}
where $\delta^{||}_{\alpha\beta}=\mathrm{diag}(0,0,1,1)$ - Kronecker symbol in the longitudinal space.
Form factors $F^l(p,k)$ have the following representation
\begin{multline}
F^l(p,k)=
\frac{2\sqrt{B}}{\pi}\sum_f q_f \mathrm{Tr}_{\hat{n}} 
|\hat{n}|^3\int^{\infty}_{0} \frac{ds_{1}ds_{2}ds_{3}}{\left(t_1+t_2+t_3\right)}
\frac{\xi_1\xi_2\xi_3 }{\left(\xi_1+\xi_2+\xi_3+\xi_1\xi_2\xi_3\right)} ~\mathcal{P}^l(s_1,s_2,s_3)
\nonumber \\
\times 
\exp\left\{ -p^{2}_{||}\phi_{1}(s_{1},s_{2},s_{3})-p_{||}k_{||}\phi_{2}(s_{1},s_{2},s_{3})-k^{2}_{||}\phi_{3}(s_{1},s_{2},s_{3})
\right. \nonumber \\ 
\left.
-p^{2}_{\perp}\phi_{4}(s_{1},s_{2},s_{3})-p_{\perp}k_{\perp}\phi_{5}(s_{1},s_{2},s_{3})-k^{2}_{\perp}\phi_{6}(s_{1},s_{2},s_{3})-m^{2}_f(s_{1}+s_{2}+s_{3})\right\}, \nonumber 
\end{multline}
\begin{gather}
t_j=B\left|\hat{n}\right|s_j, \ \ \xi_j=\frac{2t_j}{t_j\coth(t_j)+1}
\nonumber \\
\phi_1=\frac{t_1\left(t_2+t_3\right)}{t_1+t_2+t_3}, ~  
\phi_2=\frac{2t_1t_2}{t_1+t_2+t_3},~  
\phi_3=\frac{t_2\left(t_1+t_3\right)}{t_1+t_2+t_3},   
\nonumber \\
\phi_4= \frac{\xi_3\left(\xi_1+\xi_2\right)}{\xi_1+\xi_2+\xi_3+\xi_1\xi_2\xi_3}, \ 
\phi_5= \frac{2\xi_2\xi_3}{\xi_1+\xi_2+\xi_3+\xi_1\xi_2\xi_3}, ~ 
\phi_6= \frac{\xi_2\left(\xi_1+\xi_3\right)}{\xi_1+\xi_2+\xi_3+\xi_1\xi_2\xi_3}, \nonumber
\end{gather}

where $\mathcal{P}^l(s_1,s_2,s_3)$ - are the rational functions 
\begin{gather}
\mathcal{P}^1(s_1,s_2,s_3)= 
\frac{ 32\xi_1^2 X \big[ \coth(t_1)(C_{12}+4) +4D_{23} \big] 
\big[ X Y_{13}(Y_{23} +\xi_3) + Y_{13}^2Y_{23} + X^2\xi_3 \big] }
{Y_{13}\left(2Y + X\right)^3 }, \nonumber 
\end{gather}
\begin{multline}
\mathcal{P}^2(s_1,s_2,s_3)= 
\Big( 16\xi_1 X\left[ \coth(t_1)\left(C_{12}+4\right) +4D_{23} \right] \nonumber \\ 
\left[ X \left(2\xi_1Y_{23} + \xi_3 \left(Y_{23}+\xi_3 \right)\right) +   Y_{13}\left( \xi_1\left(Y_{23}+\xi_2 \right) + \xi_3Y_{23} \right) + X^2\xi_3 \right] \Big) \Big/ \left(Y_{13}\left(Y + X\right)^3 \right), 
\nonumber 
\end{multline}
\begin{multline}
\mathcal{P}^3(s_1,s_2,s_3)=
\Big( 8\xi_1 X \Big[ X \Big(\coth(t_1)\big[2\xi_1(C_{23}\xi_2+4\xi_2+\xi_3 ) + \nonumber \\ 
\xi_3(-2C_{23}\xi_2-7\xi_2+2\xi_3 ) \big] + 
D_{23} \big[2\xi_1(4\xi_2+\xi_3) + \xi_3(2\xi_3-7\xi_2) \big] \Big) + \nonumber \\
Y_{13}\Big(\coth(t_1)\big[\xi_1(2C_{23}\xi_2+8\xi_2+\xi_3) -Y_{23}(2C_{23}\xi_2+8\xi_2-\xi_3)\big]  + \nonumber \\
D_{23}\big[\xi_1(8\xi_2+\xi_3)-8\xi_2^2-7X_{23}+\xi_3^2 \big] \Big) + X^2\xi_3D \Big] \Big) \Big/ \left(Y_{13}\left(Y + X\right)^3 \right), \nonumber 
\end{multline}
\begin{multline}
\mathcal{P}^4(s_1,s_2,s_3)=
-\Big( 8 X\xi_2 \Big[ \Big( \coth(t_1) \big[ -2\xi_1^2 (C_{23}+4)+2X_{12}(C_{23}+4)- \nonumber \\ 
X_{13}(2C_{23}+7)+\xi_3Y_{23} \big] -  
D_{23} \big[ 8\xi_1^2+\xi_1(7\xi_3-8\xi_2)-\xi_3Y_{23} \big] \Big) + \nonumber \\
 X \big( \coth(t_1)[\xi_3-2\xi_1(C_{23}+4)]-D_{23}[8\xi_1-\xi_3] \big) \Big] \Big) \Big/ \left(\left(Y + X\right)^3 \right) , \nonumber
\end{multline}
\begin{multline}
\mathcal{P}^5(s_1,s_2,s_3)= 
\Big(8\xi_1 X \Big[ X \big(2\coth(t_1)\xi_1[\xi_2(C_{23}+4)+\xi_3(2C_{23}+7)]+ \nonumber \\
\coth(t_1)\xi_3[2\xi_3(2C_{23}+7)-\xi_2] +  
D_{23}(8X_{12}+14X_{13}-X_{23}+14\xi_3^2) \big) + \nonumber \\
Y_{13} \big(\coth(t_1)\xi_1[2\xi_2(C_{23}+4)+\xi_3(2C_{23}+7)] - \nonumber \\ \coth(t_1)Y_{23}[2\xi_2(C_{23}+4)-\xi_3(2C_{23}+7)] + D_{23}(8X_{12}+7X_{13}-8\xi_2^2-X_{23}+7\xi_3^2) \big) + \nonumber \\
X^2\xi_3[\coth(t_1)(2C_{23}+7)+7D_{23}] \Big] \Big) \Big/ \left(Y_{13}\left(Y + X\right)^3 \right), \nonumber 
\end{multline}
\begin{multline}
\mathcal{P}^6(s_1,s_2,s_3)=
-\Big( 8X\xi_2  \big( \coth(t_1)[-2\xi_1^2(C_{23}+4)+2X_{12}(C_{23}+4) + \nonumber \\
\xi_3(2C_{23}+7)Y_{23}-X_{13}] -  
D_{23}[8\xi_1^2+\xi_1(\xi_3-8\xi_2)-7\xi_3Y_{23}] \big) + \nonumber \\
X \big[-2\coth(t_1)\xi_1(C_{23}+4)+\coth(t_1)\xi_3(2C_{23}+7)- 
D_{23}(8\xi_1-7\xi_3) \big] \Big) \Big/ \left(\left(Y + X\right)^3 \right), \nonumber
\end{multline}
\begin{gather}
\mathcal{P}^7(s_1,s_2,s_3)=
- \frac{ 16 X\xi_2 \big[ \coth(t_1)(C_{23}+4)+4D_{23} \big]   
\big[ \xi_1^2(Y_{23}+\xi_2)+\xi_3Y_{23}  + 
X\xi_3 \big] }{ \left(Y + X\right)^3 }, \nonumber
\end{gather}
\begin{gather}
\mathcal{P}^8(s_1,s_2,s_3)=
-\frac{ 32 X\xi_2 Y_{13} \big(\coth(t_1)(C_{23}+4)+4D_{23} \big) }{\left(Y + X\right)^3 }, \nonumber
\end{gather}
\begin{gather}
\mathcal{P}^9(s_1,s_2,s_3)=
- \frac{ \xi_1 \big[\coth(t_1)(C_{23}+4)+4D_{23} \big] \big[X Y_{13}(Y_{23}+\xi_3) + Y_{13}^2Y_{23} + X^2\xi_3 \big] }{ Y_{13}^2\left(Y + X\right)^3  }, \nonumber
\end{gather}
\begin{multline}
\mathcal{P}^{10}(s_1,s_2,s_3)=
-\Big( 8X\xi_1\big(Y_{13}+X\big) \big[  \big( \coth(t_1)[\xi_1(2C_{23}\xi_2+8\xi_2+\xi_3)+\xi_3Y_{23}] + \nonumber \\
D_{23}[\xi_1(8\xi_2+\xi_3)+\xi_3Y_{23}] \big) +  X\xi_3Y \big] \Big) \Big/ \left(Y_{13} \left(Y + X\right)^3 \right), \nonumber   
\end{multline}
\begin{multline}
\mathcal{P}^{11}(s_1,s_2,s_3)= \nonumber \\
- \Big( 8X\xi_2 \big[  \big( \coth(t_1) [\xi_1^2(C_{23}+4)+X_{13}(2C_{23}+7)+\xi_3(C_{23}+3)Y_{23}] + 
D_{23}[4\xi_1^2+7X_{13}+3\xi_3Y_{23}] \big) + \nonumber \\
X \big( \coth(t_1)\xi_1(C_{23}+4)+\coth(t_1)\xi_3(C_{23}+3)+D_{23}(4\xi_1+3\xi_3) \big) \big] \Big) \Big/ \left(\left(Y + X\right)^3 \right), \nonumber  
\end{multline}
\begin{multline}
\mathcal{P}^{12}(s_1,s_2,s_3)= 
\Big( 4X \big[  Y_{13}[\coth(t_1)(\xi_1(4C_{23}+15)-\xi_2-\xi_3 + D_{23}(15\xi_1-\xi_2-\xi_3)] - \nonumber \\
 X [\coth(t_1)(\xi_3-\xi_1(4C_{23}+15)) - D_{23}(15\xi_1-\xi_3)] \big] \Big) \Big/ \left(Y_{13}\left(Y + X\right)^3 \right) , \nonumber
\end{multline}
\begin{multline}
\mathcal{P}^{13}(s_1,s_2,s_3)= 
-\frac{ 8X\xi_1  \big[ \coth(t_1)(C_{23}+4)+4D_{23} \big] \big[
XY_{13}(Y_{23}+\xi_3) + ^4Y_{13}^2Y_{23}+^4X\xi_3 \big] }{Y_{13}\left(Y + X\right)^3  } , \nonumber
\end{multline}
\begin{multline}
\mathcal{P}^{14}(s_1,s_2,s_3)= 
-\Big( 8 X\xi_1  \big(Y_{13}+X \big) \big( \coth(t_1)[2X_{12}(C_{23}+4)+X_{13}(2C_{23}+7)+\xi_3(2C_{23}+7)Y_{23}]+ 
\nonumber \\
D_{23}[\xi_1(8\xi_2+7\xi_3)+7\xi_3Y_{23}] + X\xi_3[\coth(t_1)(2C_{23}+7)+7D_{23}] \big) \Big) \Big/ \left(Y_{13}\left(Y + X\right)^3 \right) , \nonumber
\end{multline}
\begin{multline}
\mathcal{P}^{15}(s_1,s_2,s_3)= \nonumber \\
- \Big( 8X\xi_2 \big[ \big( \coth(t_1) [\xi_1^2(C_{23}+4)-\xi_3(C_{23}+3)Y_{23}+X_{13}] + 
D_{23}[4\xi_1^2+X_{13}-3\xi_3Y_{23}] \big) + \nonumber \\
X \big( \coth(t_1)\xi_1(C_{23}+4)-\coth(t_1)\xi_3(C_{23}+3)+D_{23}(4\xi_1-3\xi_3) \big) \big] \Big) \Big/ \left(\left(Y + X\right)^3 \right), \nonumber  
\end{multline}
\begin{multline}
\mathcal{P}^{16}(s_1,s_2,s_3)= 
\Big( 4X  \big[  Y_{13}[\coth(t_1)(\xi_1(2C_{23}+9)-  (2C_{23}+7)Y_{23}) + D_{23}(9\xi_1-7Y_{23})] - \nonumber \\
X [-\coth(t_1)\xi_1(2C_{23}+9)+\coth(t_1)\xi_3(2C_{23}+7) - D_{23}(9\xi_1-7\xi_3)] \big] \Big) \Big/ \left(Y_{13}\left(Y + X\right)^3 \right) , \nonumber
\end{multline}
\begin{multline}
\mathcal{P}^{17}(s_1,s_2,s_3)= 
-\frac{  8X\xi_1\big(4D_{23} + \coth(t_1)(4 + C_{23})\big)\big(X^2\xi_1 + Y_{13}^2Y_{23} + XY_{13}(Y_{23} + \xi_3) \big) }{BY_{13}^2(X + Y)^3}, \nonumber
\end{multline}
\begin{gather}
\mathcal{P}^{18}(s_1,s_2,s_3)=
- \frac{16X\xi_2 \big[ \coth(t_1)(C_{23}+4)+4D_{23} \big]  
\big[Y_{13}Y_{23} +X\xi_3 \big]}{Y_{13}\left(Y + X\right)^3 }, \nonumber
\end{gather}
\begin{gather}
\mathcal{P}^{19}(s_1,s_2,s_3)=
\frac{ 8X\xi_2 \big(\coth(t_1)(C_{23}+4)+4D_{23} \big) \big(\xi_1^2+2\xi_1Y_{23}+\xi_3Y_{23} + XY_{13} \big) }{\left(Y + X\right)^3 }, \nonumber
\end{gather}
\begin{gather}
\mathcal{P}^{20}(s_1,s_2,s_3)=
\frac{8 \big(\coth(t_1) (C_{23}+4)+4D_{23}\big) \big(Y_{13}(\xi_1-Y_{23})+ X(\xi_1-\xi_3)\big)}{(Y_{13} \left(Y + X\right)^2)}, \nonumber
\end{gather}
\begin{multline}
\mathcal{P}^{21}(s_1,s_2,s_3)= 
-\Big( 8X_{12}\xi_1 \big[X\xi_3\big(\coth(t_1)(2\xi_1(C_{23}+3)+2C_{23}\xi_3-\xi_2+6\xi_3) + D_{23}(6\xi_1-\xi_2+6\xi_3)\big)+ \nonumber \\ 
\xi_3Y_{13}\big(\coth(t_1)(\xi_3(C_{23}+3)Y_{13}-\xi_2^2(C_{23}+4)-X_{23}) - D_{23}(-3\xi_3Y_{13}+4\xi_2^2+ 
X_{23})\big)+ \nonumber \\ 
X^2\xi_3^2(\coth(t_1)(C_{23}+3)+3D_{23}) \big] \Big) \Big/ \left(Y_{13}\left(Y + X\right)^3 \right), \nonumber
\end{multline}
\begin{multline}
\mathcal{P}^{22}(s_1,s_2,s_3)= 
\Big( 8 X\xi_2   \big( \coth(t_1)[2X_{12}(C_{23}+4)+X_{13}(2C_{23}+7)+\xi_3(2C_{23}+7)Y_{23}]+ 
\nonumber \\
D_{23}[\xi_1(8\xi_2+7\xi_3)+7\xi_3Y_{23}] + X\xi_3[\coth(t_1)(2C_{23}+7)+7D_{23}] \big) \Big) \Big/ \left(Y_{13}\left(Y + X\right)^3 \right) , \nonumber
\end{multline}
\begin{gather}
\mathcal{P}^{23}(s_1,s_2,s_3)=-\frac{1}{4}\mathcal{P}^{8}(s_1,s_2,s_3), \nonumber
\end{gather}
\begin{multline}
\mathcal{P}^{24}(s_1,s_2,s_3)=
\Big( 4 X\big[ \coth(t_1)(2C_{23}\xi_1-2C_{23}\xi_2+2C_{23}\xi_3+7\xi_1-9\xi_2+7\xi_3) + D_{23}(7\xi_1-9\xi_2+7\xi_3) + 
\nonumber \\
X(\coth(t_1)(2C_{23}+7)+7D_{23}) \big] \Big) \Big/ \left(Y + X\right)^2, \nonumber
\end{multline} 
\begin{multline}
\mathcal{P}^{25}(s_1,s_2,s_3)= 
-\Big( 8X_{12}\xi_1 \big[X\xi_3\big(\coth(t_1)(2\xi_1(C_{23}+3)+2C_{23}\xi_2+2C_{23}\xi_3+7\xi_2+6\xi_3) + \nonumber \\ 
D_{23}(6\xi_1+7\xi_2+6\xi_3)\big)+  
\xi_3Y_{13}\big(\coth(t_1)(\xi_3(C_{23}+3)Y_{13}+\xi_2^2(C_{23}+4)-X_{23}(2C_{23}+7)) + \nonumber \\ 
D_{23}(3\xi_3Y_{13}+4\xi_2^2+ 7X_{23})\big)+ 
X^2\xi_3^2(\coth(t_1)(C_{23}+3)+3D_{23}) \big] \Big) \Big/ \left(Y_{13}\left(Y + X\right)^3 \right), \nonumber
\end{multline}
\begin{multline}
\mathcal{P}^{26}(s_1,s_2,s_3)= 
\frac{ 8X \big( \coth(t_1)[\xi_1(2C_{23}\xi_2+8\xi_2+\xi_3)+\xi_3Y_{23}] + D_{23}[\xi_1(8\xi_2+\xi_3)+\xi_3Y_{23}]+X\xi_3D \big) }{ \left(Y + X\right)^3 }, \nonumber 
\end{multline}
\begin{gather}
\mathcal{P}^{27}(s_1,s_2,s_3)=-\frac{1}{4}\mathcal{P}^{8}(s_1,s_2,s_3), \nonumber
\end{gather}
\begin{multline}
\mathcal{P}^{28}(s_1,s_2,s_3)= 
\frac{4X \big(\coth(t_1)[-4C_{23}\xi_2+\xi_1-15\xi_2+\xi_3]+D_{23}[Y_{13}-15\xi_2] + XD \big) }{\left(Y + X\right)^3 }, \nonumber
\end{multline}
\begin{multline}
\mathcal{P}^{29}(s_1,s_2,s_3)=
- \frac{  \xi_1 \big[\coth(t_1)(C_{23}+4)+4D_{23} \big] \big[X Y_{13}Y_{23} 
+Y_{13}(\xi_1(Y_{23}+\xi_2)+Y_{23}^2) + X^2\xi_3 \big] }{ Y_{13}^2\left(Y + X\right)^3  }, \nonumber
\end{multline}
\begin{gather}
\mathcal{P}^{30}(s_1,s_2,s_3)=
-\frac{16  X\xi_1\xi_2 [\coth(t_1)(C_{23}+4)+4D_{23}][Y_{13}+D]}{\left(Y + X\right)^3  }, \nonumber
\end{gather}
\begin{gather}
\mathcal{P}^{31}(s_1,s_2,s_3)=-\frac{1}{4}\mathcal{P}^{8}(s_1,s_2,s_3), \nonumber
\end{gather}
\begin{gather}
\mathcal{P}^{32}(s_1,s_2,s_3)=
\frac{8 X \big(\coth(t_1)(C_{23}+4)+4D_{23}\big) \big(Y_{13}-\xi_2 +X\big) }{\left(Y + X\right)^3 }, \nonumber 
\end{gather}

where the following notations is used 
\begin{gather}
X=\xi_1\xi_2\xi_3,~ X_{ij}=\xi_i\xi_j,~ 
Y=\xi_1+\xi_2+\xi_3,~ Y_{ij}=\xi_i+\xi_j, \nonumber \\
C=\coth(t_1)\coth(t_2)\coth(t_3),~ C_{ij}=\coth(t_i)\coth(t_j), \nonumber \\
D=\coth(t_1)+\coth(t_2)+\coth(t_3),~ D_{ij}=\coth(t_i)+\coth(t_j). \nonumber 
\end{gather}

\newpage

\bibliography{Refs}

\end{document}